# Investigating the 2024 swarm–like activity offshore Kefalonia Island, aided by Machine Learning algorithms

Anagnostou, V.[1], vanagno@geo.auth.gr, ORCID: 0009-0007-9303-8399 (Corresponding author)

Papadimitriou, E.[1], ORCID: 0000-0003-3574-2787

Karakostas, V.[1], ORCID: 0000-0003-4462-7094

Bäck, T.[2], ORCID: 0000-0003-1996-0805

1. Geophysics Department, School of Geology, Aristotle University of Thessaloniki, GR54124 Thessaloniki, Greece
2. Nuclear Science and Engineering department, KTH Royal Institute of Technology, SE-100 44 Stockholm, Sweden

## ACKNOWLEDGEMENTS

The GMT mapping software (Wessel et al., 2019) and specifically its PyGMT implementation was used for figure preparation. The Matplotlib python library (Hunter, 2007) was utilized for graph and diagram construction. ObsPy (Beyreuther et al., 2010) was utilized for the majority of raw data manipulation and preprocessing. We are grateful for the valuable feedback and insights of prof. Efthimios Sokos regarding the ISOLA usage and performance.

This research is partially financially supported by the artEmis Project funded by the European Union, under Grant Agreement nr 101061712. The views and opinions expressed are, however, those of the author(s) only and do not necessarily reflect those of the European Union or European Commission–Euratom. Neither the European Union nor the granting authority can be held responsible for them.

## ABSTRACT

In March 2024, a swarm–like seismic activity occurred north of Kefalonia Island, in the central Ionian Islands area. Following a machine-learning aided workflow, we compiled an enhanced seismic catalog of 2495 low to moderate magnitude earthquakes throughout a 2–month period. Spatiotemporal analysis reveals a narrow epicentral distribution of nearly E-W alignment, approximately 5km long, much longer than the length anticipated by common scaling laws for the aftershock area extension of the stronger earthquakes that did not exceed M4.0. The findings of the study indicate that the swarm-like activity is possibly triggered by a combination of fluid movements and Coulomb stress changes. The strongest earthquakes appear beyond the diffusivity curves that are within the expected upper crust values and are possibly triggered by stress transfer by the first strong earthquake. Fluid effects rapidly diminish within the first days, while the changes in the stress field due to the combined effect of the two strongest earthquakes promote the triggering of most of the weaker earthquakes of the excitation. The findings of this study reinforce the idea of swarm-like activity initiating due to interactions between stress redistributions and fluid movements in the upper crust. The rapid employment of ML tools for the compilation of robust seismic catalogs can vastly improve our understanding of the processes that drive seismicity in highly productive areas such as the Central Ionian Islands, thus leading to improved seismic hazard assessment.

## 1. INTRODUCTION

The March 2024 swarm-like seismic excitation that took place in the step-over structures offshore North Kefalonia Island, in the central Ionian Islands area, exhibited a high earthquake productivity resulting to a large number of recordings in the stations of the HUSN (Hellenic Unified Seismic Network, 1975) with the nearest station roughly 9 Km away from the centroid of the swarm. This amount of data in a segment of the Kefalonia Transform Fault Zone (KTFZ), where only moderate magnitude earthquakes are known to have occurred, raises scientific concern for its exploitation in deciphering the dynamic and kinematic properties of the activated structures. Seismic swarms and swarm-like excitations are earthquake activity clustered in space and time, lacking any clear main shock (Scholz, 2002). Swarms have been observed in a variety of geological settings, for example, in close proximity to volcanic and geothermal areas (Mesimeri et al., 2017; Shelly & Hardebeck, 2019; Li et al., 2021; Liu et al., 2024), in extensional zones (Mesimeri et al., 2016; Fischer et al., 2014; De Barros et al., 2020), near transform faults (Lohman & McGuire, 2007; Roland & McGuire, 2009) as well as Intraplate (Su et al., 2023). The spatiotemporal evolution of seismic swarms is often thought to be governed by external aseismic processes such as aseismic slow creep, fluid flow, magma intrusion, or a combination of those (Hatch et al., 2020; Sirorattanakul et al., 2022) and could be foreshocks of a large event in some cases (Peng & Lei, 2024). Recent studies in the area have shown evidence of aseismic slip motion along February 2014 activated north Palliki fault segment (Tsironi et al., 2024), the adjacent south of the study area onshore fault. In this study, we seek to elucidate the driving mechanisms behind the recent swarm-like activity, by employing machine learning algorithms and models, capitalizing on the recent advances in the applications of AI aided data analysis in Seismology. Reliable and fast methods and techniques for detecting and analyzing enormous amounts of seismic data is essential for the monitoring of the evolution of seismicity during excitations. We compile an earthquake catalog using data from the permanent seismic station network of the HUSN (1975) for a 2-month period, starting on 24 February 2024, up until 23 April 2024. We performed the detection and phase picking of the events that occurred during this period using the Earthquake Transformer (EQT) software (Mousavi et al., 2020. We implemented earthquake and phase association for the detected events using the REAL associator (Zhang et al., 2019) and located them using the HYPOINVERSE software (Klein, 2002). The absolute location is then followed by relative location using HypoDD (Waldhauser, 2001). For the strongest earthquakes in the excitation, we calculate the moment tensor solutions using ISOLA (Sokos & Zahradník, 2013) to define the geometry and kinematics of the ruptured structures. We explore the possibility of stress triggering due to coseismic slip of the strongest earthquakes. In order to provide context to the swarm–like nature of the excitation, we also constructed diffusion curves to account for the possible triggering due to fluid effects.

## 2. SEISMOTECTONIC SETTING AND FAULT MODEL

The Kefalonia Transform Fault Zone (KTFZ) is the most seismically active area in the Mediterranean and consequently in the broader Aegean area (Fig. 1). KTFZ consists of two major right-lateral branches, the northern Lefkada branch and the southern Kefalonia branch. It is recognized as the active boundary linking the continental collision zone between Eurasia and the Adriatic microplate that extends along the western coast of Greece and Albania to the north (McKenzie, 1978) and the Hellenic arc to the south, where the Oceanic lithosphere of the Eastern Mediterranean is being subducted under the Aegean microplate (Papazachos & Comninakis, 1971). Our study area is a well-studied part of the KTFZ, that has been recognized as an extensional step over zone that connects the northern (Lefkada) with the southern (Kefalonia) branch during the aftershock activity of the 2015 Kefalonia doublet ($M_w$6.1 and $M_w$6.0, on 26 January and 2 February, respectively (Karakostas et al., 2015. Regardless of their scale, strike-slip faults commonly appear segmented as echeloned, parallel, individual fault planes, separated by step-overs (Scholz, 2002. The KTFZ stepover zone comprises smaller parallel fault segments, with a WSW-ENE strike, forming a transfer zone between the two main branches of the KTFZ. This deflection of the main fault zone azimuth forms an extensional bend or duplex, facing towards the movement direction of the main fault (Woodcock & Fischer, 1986). Such fault irregularities impede rupture and often result in rupture heterogeneities as well as rupture terminations (Scholz, 2002). This observation has been repeatedly documented by historical accounts of strong earthquakes in the area being associated either with the Lefkada or the Kefalonia branch, separately (Papazachos & Papazachou, 2003). Synchronized activation of the two branches in the past, appear to have occurred due to stress transfer between them and not due to unimpeded rupture propagation (Papadimitriou, 2002).

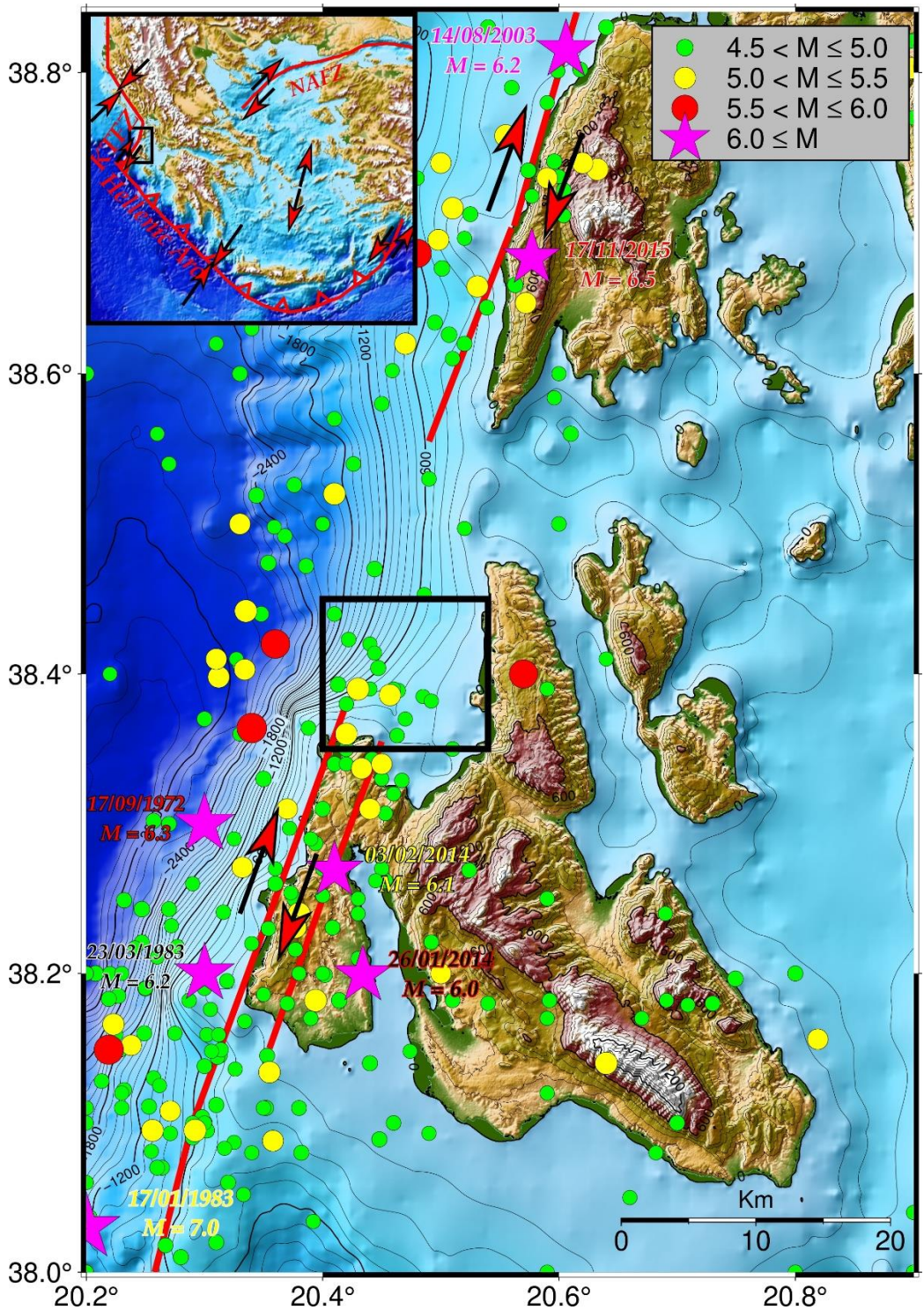

**Fig. 1** Map of the study area of the Central Ionian islands. Circles and stars of different colors and sizes denote seismicity in the region since 1970, according to the legend. The area is placed within the context of the seismotectonics of the Aegean in the map inset, along with major structures of the region, marked with red lines. The main fault segments of the KTFZ are denoted with red lines on the map of the study area. Arrows show the relative movement of the faults. Regional stations used for this study are presented in Appendix 1, Figure 7.

## 3. DATA AND METHOD

### 3.1 Data acquisition, picking and moment tensor solutions

We analyzed continuous waveform data from 12 broadband seismic stations located in the central Ionian Islands region (Fig. A1). The waveform data were provided by the EIDA services in miniSEED format in daily chunks of continuous data, resulting in a nearly 70GB raw waveform dataset. The data from each station were preprocessed by band-pass filtering within 1-45Hz frequency and resampled at a 100Hz sampling rate.

The earthquake detection and phase picking were performed by the Earthquake Transformer (EQT) software (Mousavi et al., 2020), a deep learning signal detection algorithm with an attention mechanism. The neural network is comprised of a very-deep encoder that consumes the seismic signals in the time domain and produces output information that is fed to three more encoders in a higher level, two for the phase picking (P and S) and one for the event detection. First, the algorithm performs the earthquake detection, in 60 seconds time windows, for all three components. For an event to be declared and phase picking to be performed, user defined probability thresholds are required. In our case, we fixed the threshold at 0.3 for event detection and at 0.2 for P and S phase picking, allowing picks that the algorithm considers as low probability picks in our initial pool.

The association of the seismic phases picks is then performed using the REAL software (Zhang et al., 2019) with each event requiring at least 4 pairs of P and S pickings for the catalog to be compiled. The velocity model by Haslinger et al. (1999) (Table 2, Appendix A) is used to calculate the travel-time tables for the association and for the following absolute location of the events. Absolute location of the detected and associated events is performed using the HYPOINVERSE program (Klein, 2002) followed by double-difference relocation using the HypoDD program (Waldhauser, 2001) utilizing catalog and waveform cross correlation differential times. This procedure led to a total of 47098 phase picks (22829 P and 24269 S phases) and an initial catalog of 3185 well-located earthquakes, three times more than the events included in routine analysis catalogs for the same period (24/02/2024 – 23/04/2024) and region (black rectangle shown in Figure 1). Of those 3185 earthquakes, we successfully relocated 2495 of them, with the location and relocation process being described in depth in Appendix A. We then estimate the magnitudes of the detected and relocated earthquakes, following the International Association of Seismology and Physics of the Earth's Interior (IASPEI) guidelines (Bormann & Dewey, 2014) and using the formula proposed by Scordilis et al. (2016) :

$$M_L = \log_{10}(A) + 1.2328 \log_{10}(D) + 0.0031D + 3.1465 \qquad (1)$$

Where A is the horizontal displacement maximum amplitude (mm) and D the epicentral distance (km). The detailed process of magnitude estimation and robustness assessment is detailed in Appendix B. For the strongest earthquakes, we calculate moment tensor solutions and moment magnitudes, employing the ISOLA software package for waveform modeling (Sokos & Zahradník, 2013). We modeled waveforms of earthquakes recorded by at least 4 stations each at epicentral distances of 60-200 km, using the iterative deconvolution inversion method of Kikuchi & Kanamori (1991) that is implemented in the ISOLA package, along with the CSPS method (Fojtikova & Zahradník, 2014; Zahradník et al., 2015) in cases where waveforms were too noisy. We assessed the robustness of our solutions by examining a range of quality metrics provided by ISOLA, also determining the uncertainties of our solutions. The detailed inversion parameters and solutions along with the procedures followed are presented in Appendix C.

### 3.2 Static stress changes calculations

We explore the contribution of static stress changes due to the coseismic slip of the two strongest earthquakes of the excitation, to the enhancement of seismicity on the basis of the Coulomb failure criterion, using the DIS3D software (Erikson, 1986). Stress transfer between neighboring fault segments can result in either enhancement or inhibition of seismicity, in areas where Coulomb stress changes are positive or negative, respectively. Continuous tectonic loading and occasionally coseismic slip associated with the nearby fault segments lead to accumulation of stress. Rupture takes place whenever this accumulation overpasses the fault strength, with the proximity failure being quantitively expressed through the changes in Coulomb failure function ΔCFF:

$$\Delta CFF = \Delta\tau + \mu(\Delta\sigma + \Delta p) \qquad (2)$$

where $Δτ$ is the shear stress change onto the fault plane, considered positive in the direction of the fault slip, $Δσ$ is the normal to the fault stress change, considered positive for increasing tensional normal stress, $Δp$ is the pore pressure change in the rupture area, and $μ$ is the friction coefficient, commonly taking values between 0.6 and 0.8 (Harris, 1998 and references therein). Faults move toward failure when $ΔCFS$ is positive, while negative values move the fault away from failure. During the coseismic phase, we assume undrained conditions and $Δp$ is given by Rice & Cleary (1976):

$$Δp = -B \frac{Δσ_{kk}}{3} \qquad (3)$$

where B is the Skempton's coefficient (0 ≤ B < 1) and $Δσ_{kk}$ is the summation of the diagonal elements of the stress tensor. B approaches values close to zero when the pores are filled by air, while in fluid saturated rocks, B obtains values between 0.5 and 1.0, reaching values around 1.0 in fluid–saturated soils. Experimental determinations of B for rocks indicate a range between 0.5 and 0.9 for granites, sandstones and marbles (Rice and Cleary, 1976). If we assume that in the fault zone, the diagonal components of the stress tensor are equal ($Δσ_{11} = Δσ_{22} = Δσ_{33}$) so that $\frac{Δσ_{kk}}{3} = Δσ$ and the apparent coefficient of friction can be estimated as $μ' = μ(1 - B)$ (Harris, 1998), then (2) becomes:

$$ΔCFF = Δτ + μ'Δσ \qquad (4)$$

$Δσ$ and $Δτ$ are calculated according to the fault plane solution of the next earthquake in the sequence of events, whose triggering is being inspected. Values chosen for the above constants in our study, as well as geometrical parameters for the calculations are presented in the next section (4.2).

## 4. RESULTS

### 4.1 Relocation and focal mechanisms

The epicentral distribution of the relocated activity exhibits a WNW – ESE striking narrow zone, with an approximately 5 km length (Fig. 2a). It is on par with the predominantly WNW–ESE trending, strike–slip fault planes that were computed after waveform inversions. Those epicenters are located on inferred fault segments defined by moderate magnitude earthquakes that occurred in the first 24 hours after the strongest shocks, highlighting the activation of closely located, subparallel, ESE – WNW striking, 2.5 km long inferred fault patches. Seismicity often appears to branch across multiple structures and faults in large earthquakes (e.g., Hauksson et al., 1993; Waldhauser et al., 2021; Herrmann et al., 2022) but our study and other recent works (e.g., Fonzetti et al., 2024) reveal that it may also characterize sequences of moderate events. Figure 2b presents the spatiotemporal evolution of the sequence for the whole 2-month period (24/02/2024 – 23/04/2024) and with distances measured from a reference point shown as a yellow rectangle in Figure 2a. It is clear that the majority of the relocated earthquakes that occurred during our study period, including most of the strongest events (M > 3.0) are concentrated in a 4-day period, starting on 3 March 2024 up until March 7$^{th}$. After this period, seismicity gradually decreases, with few events above 2.0 occurring until the 27$^{th}$ of March. A second small burst of seismicity is observed later during April, starting on the 8$^{th}$ of April, mainly concentrated in a smaller patch to the south of the main epicentral distribution region. Epicenters in both of those areas extend in an ESE – WNW direction, in agreement with the majority of the moment tensor solutions that were calculated for this study as well as with the known seismotectonic properties of the region (Karakostas et al., 2015; Sokos et al., 2016). This style of faulting mirrors the intricate tectonics of the region, with both north and south dipping segments suggesting that fault dips alternate or form duplexes of conjugate fault segments. All computed moment tensor solutions reveal strike–slip fault segments, many with a slight reverse component, with one nodal plane striking ESE–WNW. Taking into account the uncertainties in the calculation of moment tensor solutions, especially for moderate and small earthquakes, along with the large dip angles observed in many of them, it is difficult to decisively determine the dip direction of the proposed fault planes. This uncertainty is also observed in the performed cross-sections (Figure 3). Nevertheless, these transverse structures are expected to dip antithetically, as stepovers and jogs are commonly found separated and lineated by multiple conjugate faults (Woodcock & Fischer, 1986; Scholz, 2002; Peacock et al., 2016).

Using the computed moment tensors, we performed stress inversion for a single point (0D) in order to define the details of the local stress regime. We used the MSATSI software

9                                        22/5/2025

package (Martinez-Garzon et al., 2014) which implements the SATSI stress inversion algorithm (Hardebeck & Michael, 2006). From the stress inversion results (upper right inset in Fig. 2a) we observe that the best solution for the $\sigma_1$ (red points) and $\sigma_3$ axes (blue points) exhibit gentle plunge close to the periphery of the stereogram, while the intermediate principal stress $\sigma_2$ with very high plunge values appears at and close to the center. These results are in good agreement with the known kinematics of the region (Karakostas et al., 2015), where the sub-parallel, almost W-E striking fault segments that comprise the stepover structures in the region just north of Kefalonia Island oriented at approximately 30 degrees to S1, making it well-positioned for rupture under the evaluated stress conditions thus controlling the extensional deformation.

Using the moment magnitudes we calculated during the moment tensor inversion process, we estimated that the expected length of the activated area, as calculated from common scaling laws (namely Wells & Coppersmith, 1994; Papazachos et al., 2004) is roughly equal to 1km for the strongest earthquake in our dataset ($M_w$=4.0). Nevertheless, the areas defined from the first 48 hours aftershocks are approximately 2.5 km in length, significantly longer than the area calculated from scaling laws. This implies a lower limit to rupture dimensions in the study area, namely, that a seismicity burst whatever the maximum magnitude it comprises, cannot be originated from smaller fault patches. For detailing the geometry of the activated structures, we used the ML-enhanced relocated seismicity catalog to construct strike parallel and strike normal cross sections. The surface positions of the performed cross sections, along with the spatial distribution of the seismicity is presented in Figure 3. The strike parallel section was constructed after considering the alignment of the epicentral distribution of the relocated earthquakes and the predominant strike of the fault plane solutions for the strongest earthquakes, roughly striking at an azimuth of 270° – 280°. The surface expression of this section is the line denoted as AA' in Figure 3a, along with three lines perpendicular to that (AA1', AA2' and AA3'), along with an extra strike normal section (BB1'), perpendicular to the apparent strike of the southernmost cluster of earthquakes that mostly occurred during the last days of the excitation.

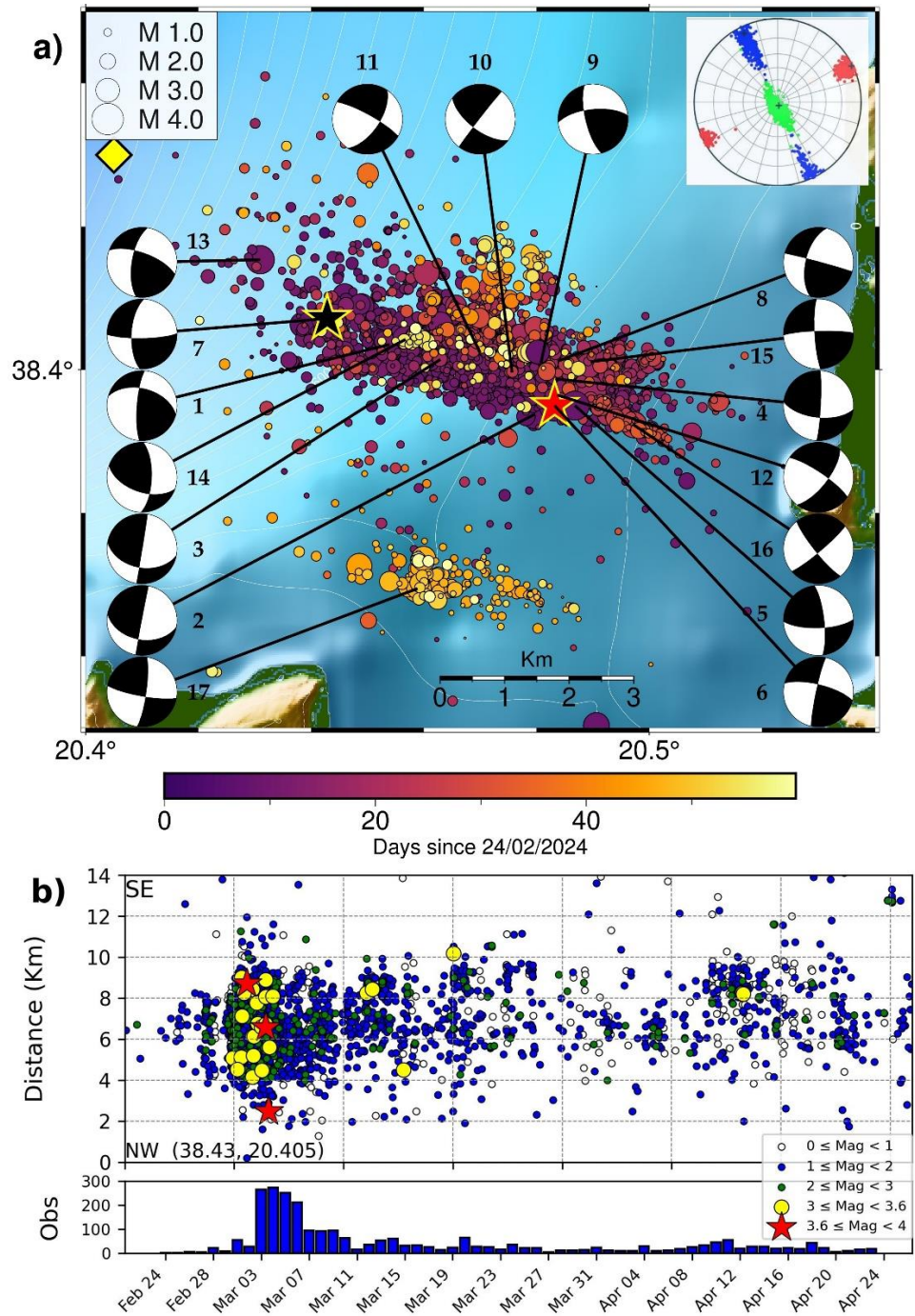

**Fig. 2** a) Relocated epicenters of the study area during the 24/02/2024 to 23/04/2024 period. The two strongest earthquakes of the sequence on 4 March 2024, 07:06:44 ($M_w$ = 4.0) and on 4 March 2024, 17:11:11, ($M_w$ =3.9) are shown as a red and black star respectively. The focal mechanisms of the 17 strongest earthquakes of the excitation are also presented, shown as equal area lower hemisphere projections. The ID number of each focal mechanism is according to table 2 in appendix C. Inset on the top right shows a stereographic plot of the three principal axes, after stress inversion of the focal mechanisms and for 500 bootstrap resamplings. Red, green and blue points denote the 95% confidence region for the $\sigma_1$, $\sigma_2$ and $\sigma_3$ principal axes respectively. Epicenters are colored according to their temporal proximity to the first day of the sequence, as shown in the colorbar and are sized according to their magnitude, following the legend shown as inset on the top left of the figure. Yellow rectangle denotes the initial measurement point for the spatiotemporal plot of the sequence b) Spatiotemporal plot, starting at 24/02/2024 and with distances measured from a point located to the Northwesternmost edge of the activated area at latitude 38.43 and longitude 20.405. Frequency of observed earthquakes per day are also presented as histograms.

As apparent from the strike parallel section (Figure 3b), seismicity for the whole of the excitation clusters within a 5-10 km depth layer, whereas the 3 strong earthquakes got focal depth of approximately 9km. The majority of the strongest (M > 3.0) earthquakes in our dataset concentrates close to this depth, in agreement with observed characteristics of the seismogenic layer and mechanical properties of the upper crust in the region (e.g., Bountzis et al., 2021). The three cross sections normal to the AA' line (A1A1', A2A2' and A3A3') are constructed at the immediate vicinity of the strongest earthquakes, starting from the northwestern edge of the activated area and proceeding towards the southeast, perpendicular to the predominant strike, each with a 2km width either side of the surface projection and for the first 24 hours after the occurrence of the strongest earthquake. The chosen time window adequately defines the dimensions of the activated structures, revealing narrow concentrations of relocated aftershocks with M>2.0 that outline the relatively high dips of the inferred slipped fault patches, consistent with the calculated focal mechanisms. Such steep dips have previously been observed during studies in the broader KTFZ region (e.g., Papadimitriou et al., 2017). The dips of the fault segments that correspond to the strongest earthquakes in our study as determined by waveform modeling, can be sufficiently constrained, steeply dipping either to the south (Figures 3c, 3d) or to the north (Figure 3e), reflecting the kinematics of the area. A persistent observation in the three cross sections is that during the first 24 hours after each strong earthquake, seismicity appears roughly constrained at a depth range of 8–10km, defined by their strongest aftershocks (M>2.0). The same depth range can be observed in the final cross section, performed for the southernmost cluster (Figure 3f). The almost vertical dip (~ 80°) of the computed focal mechanism for the strongest earthquake of the April subcluster (11-04-2024, M = 3.5) is evident in the steeply dipping focal distribution of the B1B1' cross section.

### 4.2 Static stress interactions

Previous work done in deducing the physics behind strong and moderate earthquake sequences in our study area has shown that stress changes induced by coseismic rupture play a major part in the driving forces that promote earthquake productivity (e.g. Karakostas et al., 2015; Papadimitriou et al., 2017; Kourouklas et al., 2023; Bonatis et al., 2024). We explore the contribution of static stress changes associated with the coseismic slip of the 4 March 2024, 07:06:44, $M_w$ = 4.0 and 4 March 2024, 17:11:11, $M_w$ = 3.9 earthquakes to the encouragement of seismicity based on the Coulomb failure criterion, using the DIS3D software (Ericsson, 1986). Rupture models required for the computation of ΔCFF are approximated as rectangular planes embedded in the upper crust. The geometry parameters of those planes, specifically their length and width, have been constrained by our spatiotemporal analysis, as well as the computed focal mechanisms and are presented in Table 1. Values for the scalar moment of the two earthquakes are computed during the moment tensor determination and along with the proposed fault plane dimensions were used to calculate the mean slip, by the equation:

$$M_o = \mu * S * u \qquad (5)$$

where $M_o$ is the scalar moment, $\mu$ is the shear modulus and $S$ the fault area. A homogenous elastic half space was assumed, and a value of 0.4 was chosen for the apparent coefficient of friction $\mu'$, which was found as a proper one for the central Ionian Islands area after previous work on the stress evolutionary model for the area (Papadimitriou, 2002). The shear modulus and Poisson's ratio were taken equal to 33 GPa and 0.25, respectively, in all calculations of this study. Stress changes due to the coseismic slip of the two strong earthquakes were calculated according to their faulting type.

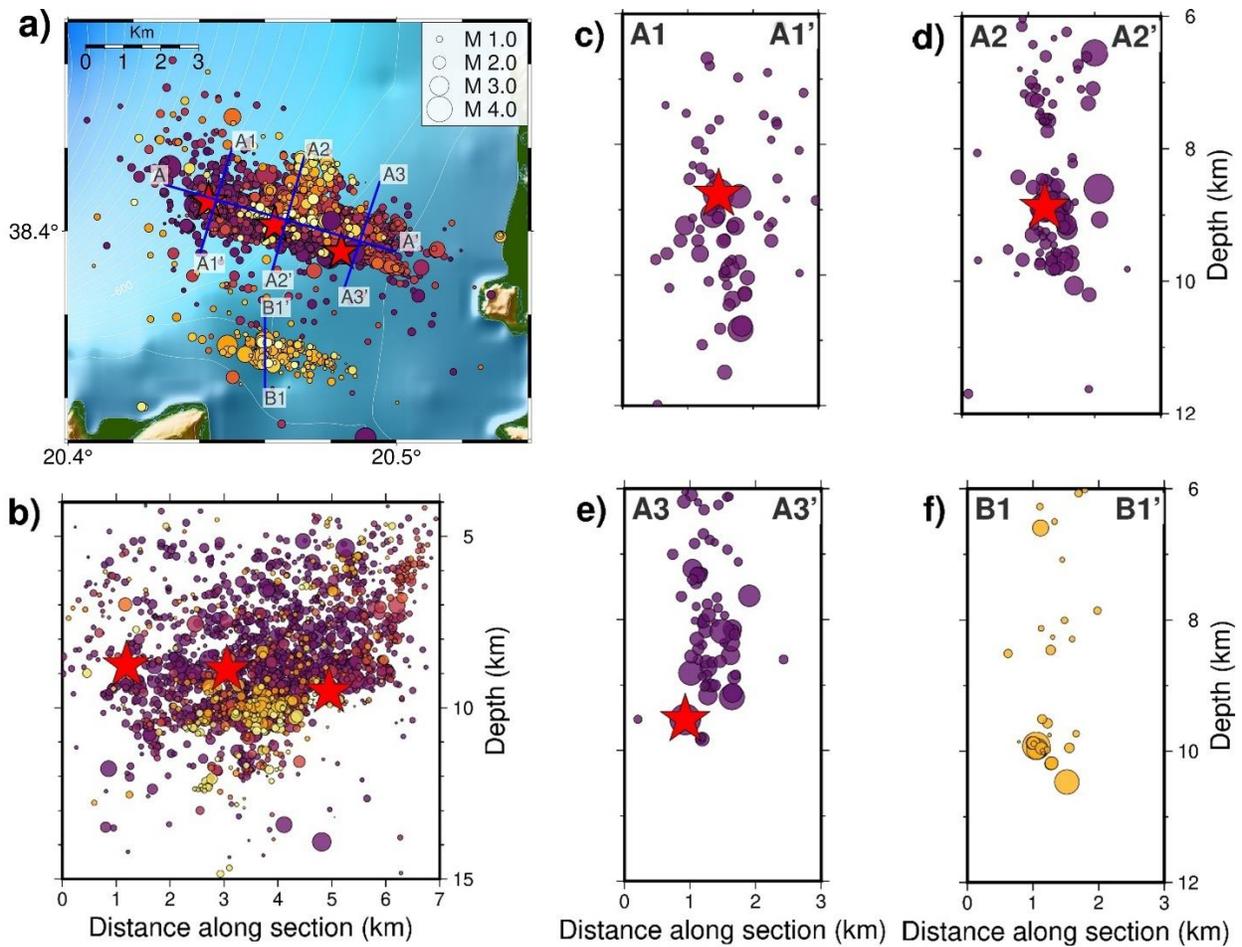

**Fig. 3** a) Spatial distribution of the seismicity and surface positions (continuous grey lines) of the performed cross sections. b) 2km wide, along strike cross section for the 2024 March – April excitation. c) 2 km wide strike normal section along the A1A1' line, comprising seismicity that occurred during the first 24 hours after the 4 March 2024, 17:11:11, $M_w$ =3.9 earthquake, denoted with a red star. d) Same as in c, but for the 3 March 2024, 13:35:09 07:06:44, $M_w$ = 3.7 earthquake, along the A2A2' line e) Same as in c but for the 3 March 2024, 07:06:44 $M_w$ = 4.0 earthquake, along the A3A3' line f) 2 km wide strike normal section along the B1B1' line, comprising seismicity that occurred during the first 24 hours after the 11-04-2024, $M_w$ = 3.5 earthquake

Figure 4a shows the Coulomb stresses calculated due to the coseismic slip of the 4 March 2024 $M_w$=4.0 earthquake, for a receiver fault with strike = 89°, dip = 80° and rake = -25° which is the faulting type of the 4 March 2024 $M_w$=3.9 earthquake. The first 48 hours shocks shown in Figure 4a include the next strong earthquake, that occurred roughly 10 hours after the first.

It is apparent that significant (>0.1 bar) stress changes due to the coseismic slip of this earthquake are limited in their spatial extent, as expected from its relatively small magnitude. While most of the 48-hour aftershocks are located in areas of negative stress changes, we observe that the second strong earthquake (shown as black star) is located in a stress enhanced area, marginally within an area where the positive Coulomb stress changes were larger than 0.1 bar. This implies that the occurrence of the second earthquake a few hours later may be encouraged by stress transfer, but this cannot be definitively stated when taking into account the inherent location uncertainties in fault modeling and therefore Coulomb stress changes calculations. Figure 4b shows the Coulomb stress changes calculated due to the coseismic slip of the second 4 March 2024, $M_w$ = 3.9 earthquake for the same earthquake faulting type. The events that occurred within the first 48 hours are also depicted, with most of them located again in areas where the Coulomb stress changes were either negative or minimally positive.

**Table 1**: Rupture models for the earthquakes included in the stress calculation model.

| Date | Time | Latitude | Longitude | L (km) | W (km) | u (cm) | $M_w$ | Focal Mechanism | | |
| --- | --- | --- | --- | --- | --- | --- | --- | --- | --- | --- |
| | | | | | | | | Strike | Dip | Rake |
| March 4th, 2024 | 07:06:44 | 38.3950 | 20.4832 | 2.5 | 2.5 | 0.5 | 4.0 | 284° | 73° | 6° |
| March 4th, 2024 | 17:11:11 | 38.4071 | 20.4428 | 2.5 | 2.5 | 0.4 | 3.9 | 89 | 80 | -25 |

To further assess and quantitatively evaluate the triggering mechanism due to the coseismic slip of the first strong earthquake, as well as the combined effect the two ruptures had on the seismicity for the rest of the study period, we calculated the Coulomb stress changes at each focus of the relocated earthquakes. Figure 5 (a) presents a histogram of these values for the earthquakes that occurred between the first and the second strong shocks, while Figure 5 (b) shows a histogram of the calculated ΔCFF due to the changes on the stress field imparted by the combined effect of both strong shocks at the focus of the earthquakes that occurred during the remaining period. In the first case, nearly 50% of the earthquakes obtained negative stress changes, implying that those earthquakes occurred due to the rupture of the causative fault of the first strong earthquake, as it is apparent that the epicenters are located on the inferred fault area (Figure 4a). In the second case (Figure 5b) we observe that more than 60% of the earthquakes of the excitation occurring after the second strong shock, obtain positive ΔCFF values.

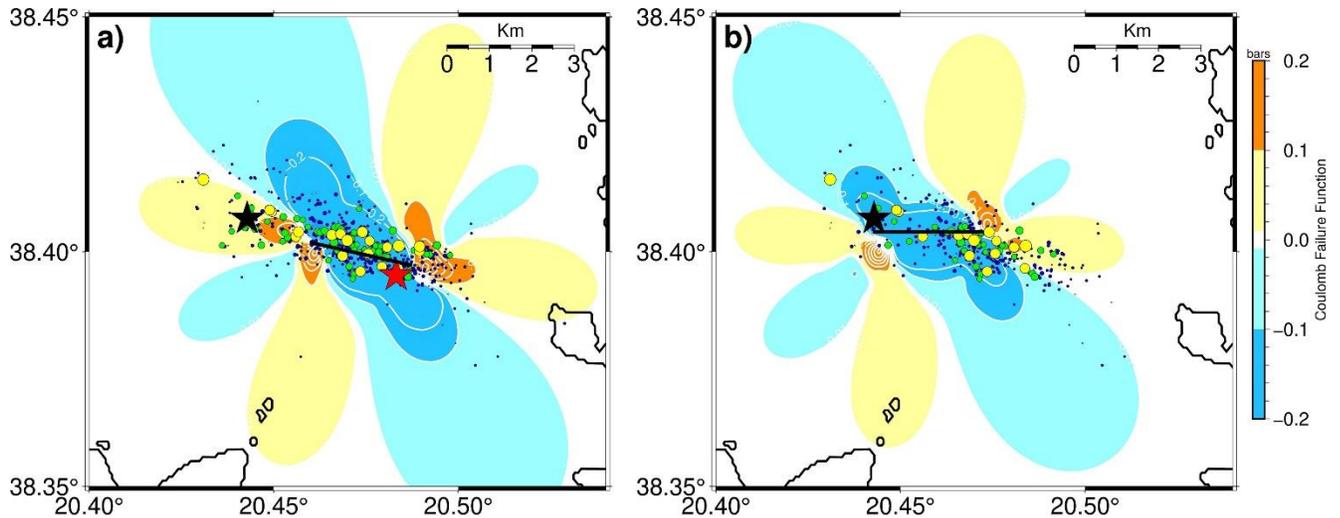

**Fig. 4** a) Coulomb stress changes due to the coseismic slip of the Mw=4.0 earthquake (red star), calculated according to the Mw=3.9earthquake faulting type and at a depth of 8km. The black star denotes the second (Mw = 3.9) strong earthquake, occurring 10 hours after the first and placed in a region of high, positive ΔCFF. Colored circles correspond to the epicenters of the earthquakes that occurred 48 hours after the first shock. The thick black line denotes the inferred fault patch of the Mw = 4.0 earthquake b) Coulomb stress changes due to the coseismic slip of the Mw=3.9 earthquake (black star), calculated according to the earthquake faulting type. Colored circles correspond to the epicenters of the earthquakes that occurred 48 hours after the second (Mw = 3.9) shock. The thick black line denotes the inferred fault patch of the second earthquake. Stress changes for both panels are according to the color scalebar on the right (in bars).

The above percentages may be seen by visual inspection of the normalized cumulative curve in Figure 5, where the line yielding point occurs at the corresponding percentages. While these numbers provide strong evidence that the combined effect the two strongest earthquakes had on the stress field promoted the triggering of a large percentage of the shocks that followed the first 10 high-productivity hours of the sequence, there is a relatively large percentage of earthquakes occurring in areas of positive, yet low (below 0.1 bar) ΔCFF values. This fact drives us to further investigate other physical mechanisms that may have affected the evolution of the excitation.

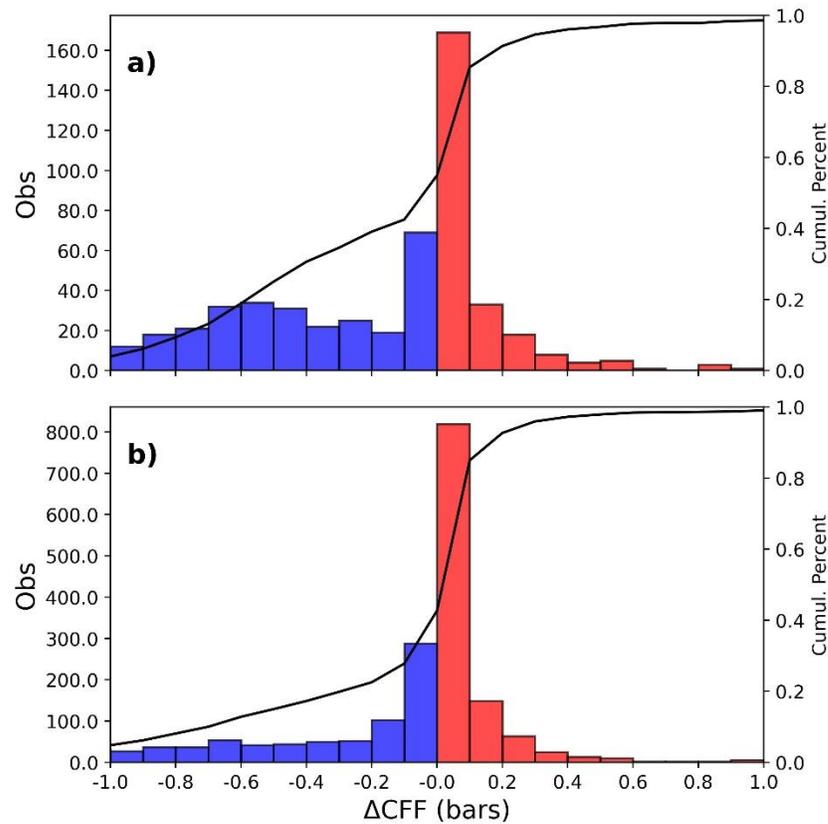

**Fig. 5** Histograms of the Coulomb stress changes at the focal depths: a) of the earthquakes that occurred between the first 04/03/2024, $M_w$=4.0 earthquake and the second 04/03/2024, $M_w$=3.9 earthquake, according to the fault plane solution of the second earthquake and due to the coseismic slip of the first earthquake b) of the earthquakes that occurred after the second 04/03/2024, $M_w$=3.9 earthquake, according to the fault plane solution of the same earthquake and due to the coseismic slip of both fault segments

### 4.3 Fluid effects on the seismicity evolution

We assess the possible triggering effect that pore pressure perturbations may have in this excitation. The spatiotemporal diffusive plot of the sequence (Fig. 6a) is constructed with reference point (both spatially and temporally) the 4 March 2024 $M_w$ = 4.0, at the southeastern edge of the activated area. During the highest productivity period of the sequence, the first earthquakes were detected close to the chosen reference point, while some earthquakes of moderate magnitude occurred on the opposite side of the activated region. Figure 6a contains the majority of the catalog, namely the shocks that occurred during the first 50 days. The diagram shows an initial rapid activation of the fault system within the first 72 hours and a progressive drop in seismicity, both in magnitudes and occurrence frequency. The strongest earthquakes that occur beyond the first 72 hours, appear to follow the triggering fronts that correspond to low diffusivity coefficients (0.1 – 0.3 m$^2$/s). For the same period after the occurrence of the two strongest shocks, we observed that a large percentage of the seismicity can be attributed to stress changes imparted by the combined slip of the two shocks. Taking a closer look in the first 72 hours after the first strong shock and in its spatial proximity (Fig. 6b), we observe that the triggering front that corresponds to the D=28 m²/s diffusivity coefficient effectively reproduces the spatiotemporal diffusion of the activity within this time window. Taking into account the inherent uncertainties in Coulomb stress analyses calculations and the fact that the second largest earthquake (Mw = 3.9) appears marginally within areas of significant positive stress changes, we can attribute its occurrence as well as the occurrence of the earthquakes within this high productivity time period to fluid diffusion. Even though classical understandings of expected diffusivity coefficient values in the upper crust are of lower values (between 0.01 and 10 m$^2$/s) (Talwani & Acree, 1985; Scholz, 2002), recent experimental studies (Lei, 2024) as well as swarm studies (Antonioli et al., 2005; Lei et al., 2024; Liu et al., 2024) have shown that D can reach values as high as 100 m$^2$/s.

From the above, there is strong evidence that fluid diffusion may be the primary driving factor behind this swarm-like activity, with the observed stress changes due to the coseismic slip of the two strongest earthquakes playing a secondary role in the cascade triggering of the earthquakes that occurred after the first highly-productive days of the sequence. This interaction between coulomb stress triggering and fluid diffusion is a previously observed phenomenon (e.g., Piombo et al., 2005; Mesimeri et al., 2017; Sirorattanakul et al., 2022). Porous solids deform in response to stress changes, changing the pressure of fluids within the pores and possibly promoting their flow while swarm seismicity commonly expands outwardly from a reference point (e.g., an injection point or a strong earthquake), in a way consistent with a diffusive process (Shapiro et al., 1997; Hainzl, 2004; Shapiro & Dinske, 2009; Chen et al., 2012). This is also clearly evidenced within the first 72 hours of the excitation, where seismicity apparently migrates towards the northwestern edge of the activated area, up to a lateral distance of roughly 4.5km from the epicenter of the reference earthquake, to be then comprised within a roughly 2km long area close to the initially activated area.

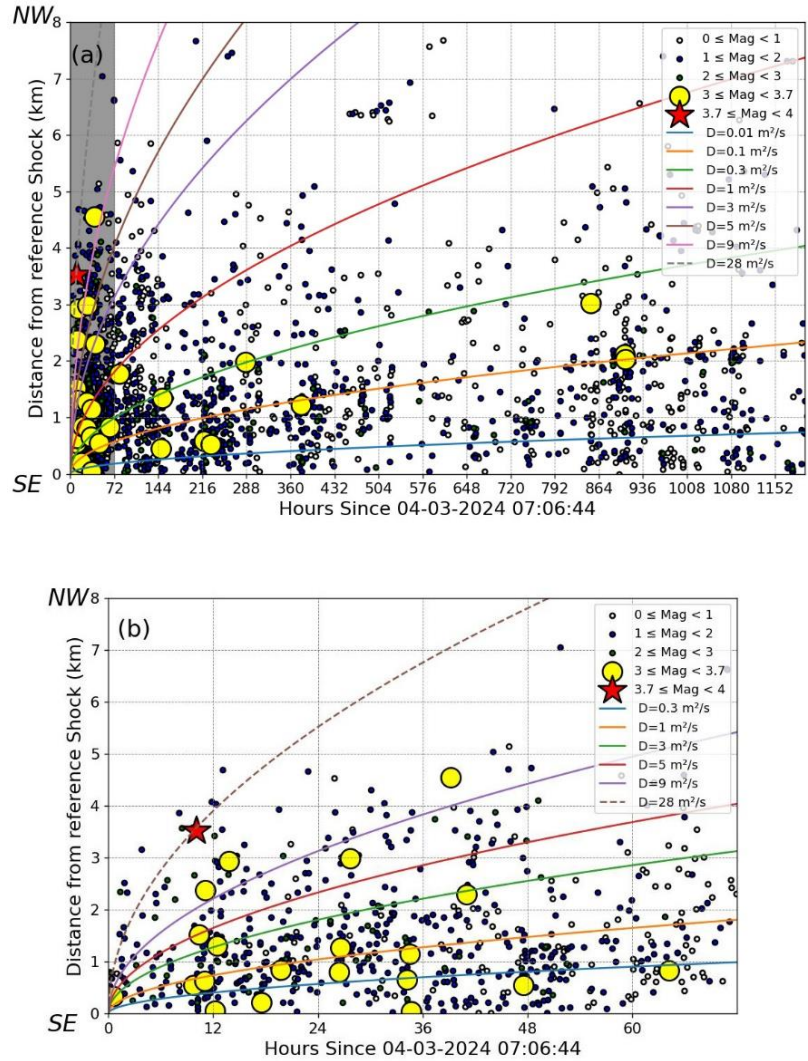

**Fig. 6** Pore pressure diffusion as a driver for the evolution of the excitation. a) Diffusion plot with the distance from the origin of the 4 March 2024 $M_w$=4.0 earthquake versus the occurrence time up until 23/04/2024. Colored lines denote the trigger front curves, each corresponding to a different diffusivity coefficient. The fault system appears to activate rapidly in the first 72 hours of the excitation, denoted with a gray box. b): Detailed view of the grey area for the first 72 hours of the excitation. The triggering front that corresponds to the D =28 m²/s diffusivity coefficient appears to drive the seismicity within this high-productivity period of the sequence, along with the occurrence of the second strongest $M_w$ = 3.9 earthquake.

## 5. DISCUSSION AND CONCLUSIONS

Seismic activity during swarm sequences can potentially shake the activated region for a long time, raising significant concerns and making it crucial to infer potential evolutionary scenario. We attempt to deduce such scenario for the seismic excitation that took place in the offshore area north of Kefalonia Island, during March and April 2024. Considering the study area and its individual seismotectonic properties as a natural laboratory, we seek to test and verify the potential ability to derive rapid and high–quality data by applying Machine Learning tools and methods. We successfully compiled an enhanced seismic catalog and subsequently refined it with robust location and relocation techniques, an extremely valuable prospect in studying the seismicity evolution in very seismically productive areas, such as the central Ionian Islands. Our ML catalogue yields a better constraint on the spatiotemporal characteristics of the excitation when compared to the routine catalogue, presenting a significant lowering of the mean RMS (0.24 vs 0.14 seconds) whereas also including nearly 3 times more events.

Utilizing the enhanced catalog, we attempt to shed light on the interactions between the structures. Focusing on two of the strongest earthquakes, we found that the shocks that occur in the first 24 hours during the highest productivity period of the excitation, align on sub-parallel, ESE – WNW striking fault segments that activate within few hours of one another and define the activated area. We found that seismicity branches on closely located fault segments, despite the moderate magnitudes that we observed. This branching is often obscured behind the unrefined epicentral distributions as located by standard routine location procedures, leading to loss of valuable insight in regional seismotectonics. Our relocated catalogue refines and enhances these patterns, indicating that branching and remote activation of small fault segments is present in smaller scales as well. Our study region exhibits a significantly larger activated area and apparent fault segment length, despite the values expected by common scaling laws, implying a lower limit to rupture dimensions. The epicentral distribution along with the computed focal mechanisms for 17 of the strongest shocks in our dataset using waveform modeling, provide further support for the presence of a fault system that strike obliquely to the main branches of the KTFZ, forming a strike slip duplex of conjugate faults that separates the two.

We observe a rapid activation with a subsequent drop in seismicity rate within the first 12 days after the first strong earthquake. Triggering of the second strongest shock appears to be attributed to Coulomb stress changes due to the coseismic slip of the first earthquake. The combined effect the two shocks had on Coulomb stress changes in the area supports the triggering of a large percentage of the aftershocks. We found that fluid related processes and movements may secondarily drive the activation of smaller segments and thus the seismicity that is beyond the triggering capabilities of the ΔCFF. Moderate values of the diffusivity coefficient can sufficiently explain most of the seismicity occurring during the first 72 hours, whereas lower values can explain some of the strongest earthquakes occurring a month or later after the initiation of this swarm-like sequence.

The physics driving the activation and evolution of swarm and swarm-like excitations are rarely unique, often stemming from a combination of mechanisms and processes that promote them. Our study shows that a combination of triggering due to stress changes induced by the coseismic slip of the two strongest shocks in the first days of the excitation, along with perturbations inflicted by fluid movements may have driven the swarm-like sequence. The rapid employment of ML tools and algorithms and the robust location and relocation procedures is extremely valuable in tracing and studying such excitations in areas of high seismotectonic interest, where good monitoring and understanding is crucial for subsequent seismic hazard assessment.

APPENDIX A – Location and Relocation

For the detection and picking of earthquake recordings, we selected readings from stations up to the distance of ~60 km from the activated area, encompassed inside the dashed red circle (Fig. 7) and shown as inverted red triangles, along with the epicenter of the strongest shock (03-03-2024, $M_w$ = 4.0) shown as a magenta star for reference. The inverted yellow triangles depict the sites of the stations the recordings of which were used for the moment tensor inversions. For the application of the CSPS method, required clear first motion polarities were also obtained from the stations within the ~60km radius dashed circle. More details on the method can be found in Appendix C.

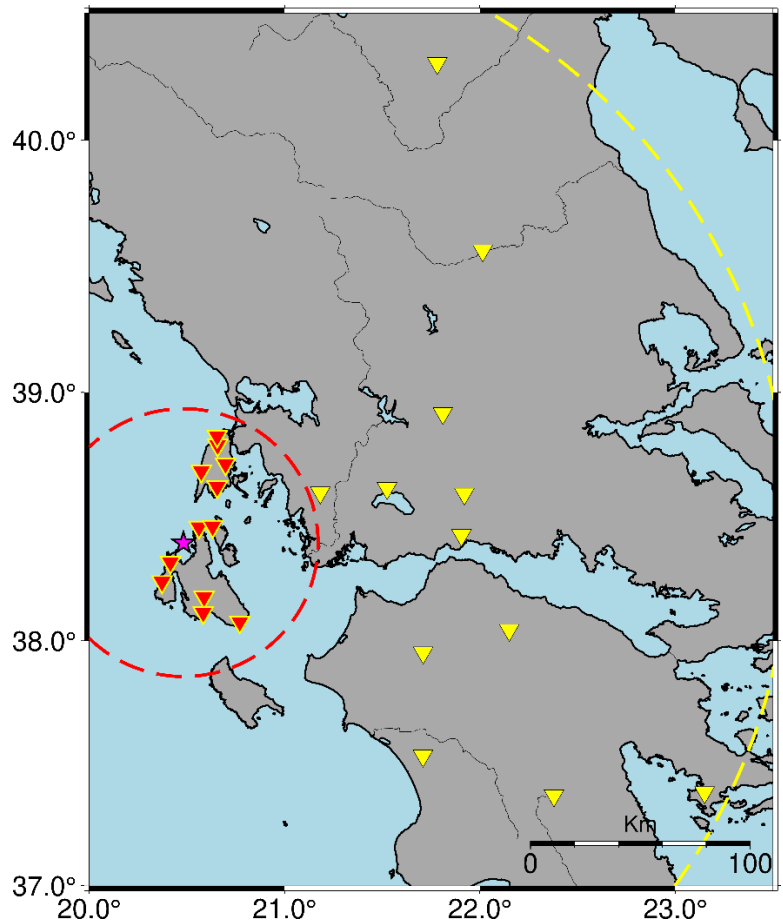

**Fig. 7** Stations of the regional Hellenic Unified Seismological Network (HUSN), the continuous recordings of which were used for earthquake detection and picking using EQT (red inverted triangles, encompassed in a red dashed circle of 60 km radius). The same stations were used to obtain clear first motion polarities, needed for the application of the CSPS moment tensor determination procedure (Appendix C). Yellow inverted

triangles, encompassed in a yellow dashed circle of nearly 270 km radius, denote stations of the HUSN the recordings of which were used for moment tensor inversions

In order to remove false positive picks and locations, we associated the initial seismic phases picked by the ML model, using the REAL algorithm (Zhang et al., 2019). We utilized a local velocity model by Hasslinger et al. (1999) (Table 1) in order to compute the theoretical travel time windows for the phase association. REAL associates a group of P and S picks with an earthquake based on their appearance within the theoretical travel time windows. A grid is constructed and based on the number of picks for each grid cell, the best location is chosen for the earthquake epicenter. In the case of multiple cells having the same number of P and S phases, the cell with the smallest travel time residual is chosen.

**Table 2**: Velocity model used for the association, location and relocation of the epicenters (Hasslinger et al., 1999)

| Depth (km) | $V_p$ (km/s) |
|---|---|
| 0.0–2.0 | 3.50 |
| 2.0–5.0 | 5.47 |
| 5.0–10.0 | 5.50 |
| 10.0–15.0 | 6.00 |
| 15.0–20.0 | 6.20 |
| 20.0–30.0 | 6.48 |
| 30.0–40.0 | 6.70 |
| 40.0–70.0 | 6.75 |
| ≥70.0 | 8.00 |

We require at least 8 seismic phases (both P and S) from at least 4 stations in order to determine a reading as an earthquake. Earthquakes with few numbers of picks, very large azimuth gaps and large travel time residuals are discarded as low quality or false events. The above constraints are chosen after multiple association tests and visual inspection of the waveforms that correspond to the lower quality associated earthquakes. The resulting associated picks are then located using the Hypoinverse program (Klein, 2002). A minimum of 4 stations and 8 phases for each event is set as a constraint for the location strategy. The located events are then further inspected and low-quality solutions and outliers are discarded from the catalog. The resulting catalog includes 2630 well-located events with an average Root-mean square error in the time domain of 0.145s±0.05(1σ), mean azimuthal gap of 180.93°±20.28° (1σ), mean depth error of 1.29km±1.08(1σ) and mean horizontal error of 0.87km±0.463 (1σ). The relatively large mean azimuth gap of the located earthquakes is attributed to the geographic location of the epicenter distribution, having no station coverage on the western side.

Next, we employ the double-difference relocation algorithm of (Waldhauser & Ellsworth, 2000) using the HypoDD program (Waldhauser, 2001) for improving relocation. Catalog differential times were considered for the relocation process, employing the conjugate gradients (LSQR) method with the appropriate damping parameter for each of the 25 performed iterations. The initial 5 iterations were performed without time and location weighting, followed by 4 sets of 5 iterations with progressively higher factors of downweighting. For the waveform cross-correlation procedure, we applied a second-order Butterworth band-pass filter of 1-10 Hz to mitigate the effects of background noise, after visual inspections on the waveform dataset. The differential times were calculated for windows of 2 seconds around the P and S phase picks, starting 0.3 seconds before the P phase arrival and 0.5s before the S phase arrival. In order to ensure that we did not include the S phase when calculating cross-correlation differential times for the P picks, in the case the time window after the P arrival time ($t_p$) exceeded the value $0.9*(t_s-t_p)$, we replaced it with this value. We applied the same constraint for the S arrival time, for a value of $0.5*(t_s-t_p)$. We discarded any pairs with a cross-correlation coefficient lower than 0.7. We then combine catalog differential times with cross–correlation differential time (Schaff & Waldhauser, 2005) for the second part of the relocation procedure, performing a total of 25 iterations. For the first 10 iterations, the cross-correlation data are downweighted in order to let the catalog differential times control the initial relocated hypocenters. Iterations 6-10 are still controlled by catalog data, but a time residual threshold as well as an event separation threshold is applied to remove outliers. In the remaining 15 iterations, we improve the location of the events whose waveforms correlate by down-weighting catalog data relative to the cross-correlation data. We allow large separation distances to still be controlled by catalog differential times while cross-correlation data constrain only event pairs with separations smaller than 3 (Iterations 11-20) and then finally 0.5 (iterations 21-25) km. Time residual threshold weights for the cross-correlation data are applied only for the last 10 iterations (16-25). After the relocation procedure, the resulting catalog contained 2495 relocated earthquakes. Since errors estimated by HypoDD using the LSQR inversion method are often underestimated (Waldhauser, 2001), we calculated the errors in horizontal and in depth, using a bootstrap resampling approach (Efron, 1982; Efron & Tibshirani, 1994). We performed 100 iterations, each with a different sample that has been modified by random perturbations of the time residuals of the original results. The error estimates are then obtained from the median absolute deviations of the bootstrap distribution of the epicentral coordinates (Leys et al., 2013). Thanks to the well contained, dense seismicity as well as the quality of raw data obtained by the dense local station network, we obtain errors in horizontal (0.130 km), depth (0.204 km) and time (0.02 s) an order of magnitude lower than those obtained during the absolute location step.

### APPENDIX B – Magnitude Estimation

Catalog magnitudes for this study were estimated by applying the formula proposed by Scordilis et al. (2016) which is widely used in routine analysis of earthquakes in the region and is suitable for local magnitude estimation:

$$M_L = \log_{10}(A) + 1.2328 \log_{10}(D) + 0.0031D + 3.1465 \quad \text{(B1)}$$

where A is the average zero-to-peak maximum amplitude for the two horizontal components (N-S and E-W), measured in millimeters, while D is the hypocentral distance (in km). We followed the procedure described by the International Association of Seismology and Physics of the Earth's Interior (IASPEI) (Bormann and Dewey, 2014) for the magnitude estimation, synthesizing Wood-Anderson recordings from the digital recordings of the stations used for the earthquake detection and picking. Simulation filters were applied on the sensor recordings for the transformation, using a 2080 sensitivity and a 0.7 damping constant, a procedure described by Uhrhammer & Collins (1990). Since one magnitude is estimated for each of the horizontal components, the final magnitude for each earthquake is the median of the two values. We then compare our magnitude estimations with the estimated magnitudes from a manual unpublished catalog for the same region and the same period (Fig. 8). Out of the 2495 relocated earthquakes of our catalog, we matched 1680 with manually picked and analyzed earthquakes, by searching for a match in origin time with a 1 second tolerance window between the two catalogs. The manual catalog magnitudes were determined using the same method as the ML catalog. We find a very good correlation between the two catalogs, with a high coefficient of determination (0.92). To improve our estimated magnitudes, we use the linear equation that describes the relationship between the magnitudes of the two catalogs to correct our values by deriving 0.15 from each.

The local Magnitude ($M_L$) and moment magnitude ($M_w$) scales are known to be equivalent for a wide range of values (Hanks & Boore, 1984; Heaton et al., 1986; Bollinger et al., 1993; Thio & Kanamori, 1995; Uhrhammer et al., 1996; Papazachos et al., 2002; Utsu, 2002; Grünthal & Wahlström, 2003; Brazier et al., 2008). With this in mind and in an attempt to further evaluate our magnitude estimations, we compare the magnitudes estimated for our ML catalog with the moment magnitudes estimated after moment tensor inversion for 15 earthquakes of the excitation.

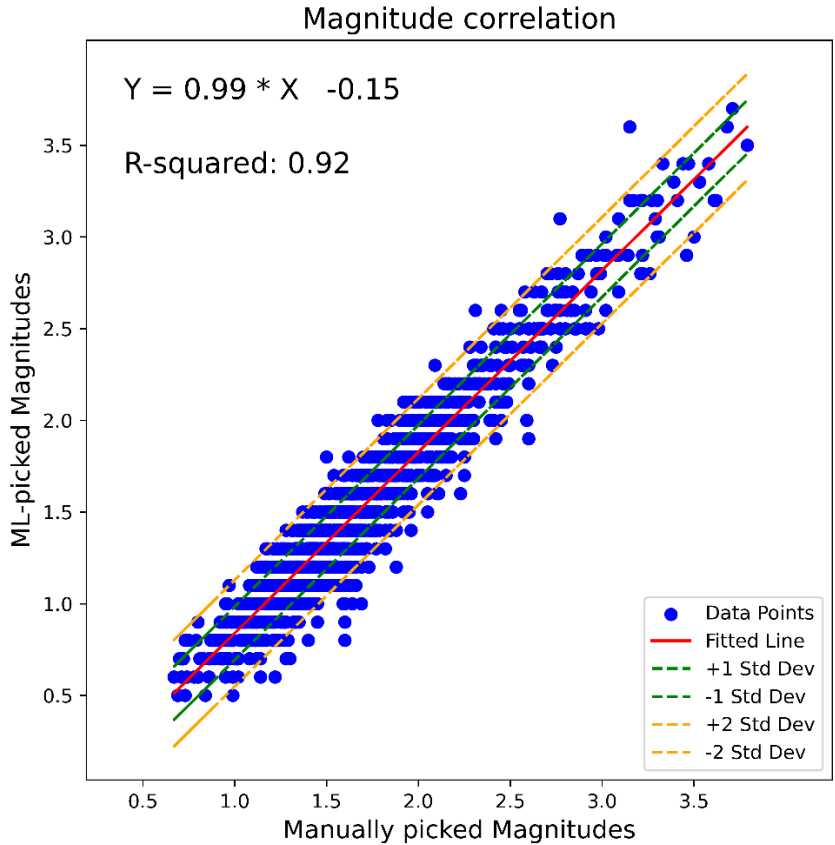

**Fig. 8** Correlation between ML-picked and manually picked local magnitudes

Figure 9 presents a direct comparison between original M_w and M_L (as estimated in the present work) of these 15 events. The relation from this comparison is:

$$M_W = 1.08 * M_L - 0.27, R^2 = 0.93$$

Applying this equation to our initially calculated local magnitudes for the same 15 earthquakes, we derive differences from 0.027 to 0.1613, with an average difference of 0.087. Those errors are within the expected errors during magnitude estimation, and as such we can consider the calculated magnitudes as reliable and practically equivalent to the calculated moment magnitudes.

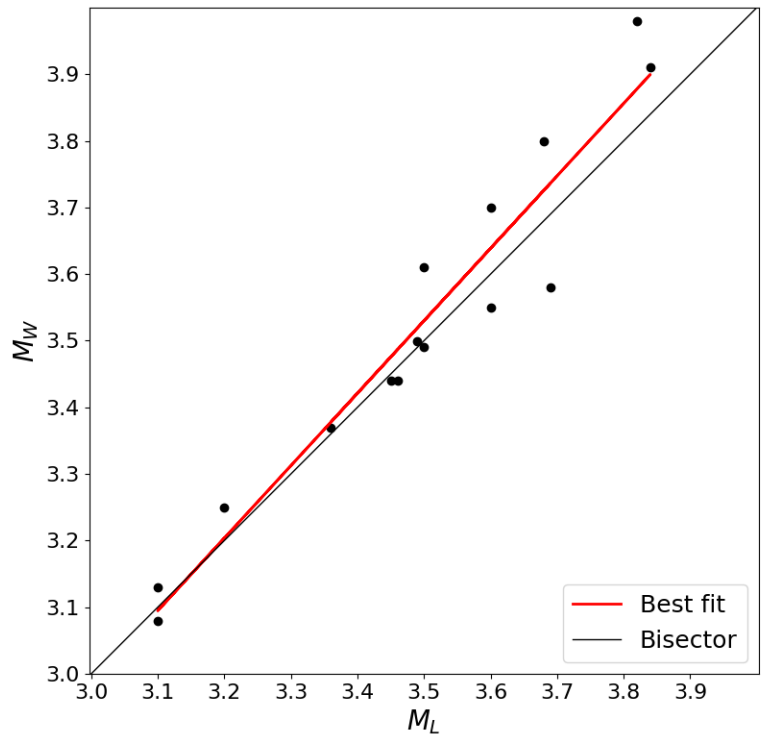

**Fig. 9** Correlation between estimated M_w and M_L

APPENDIX C – Fault Plane Solutions

We employed the ISOLA software package (Zahradník and Sokos, 2018) to calculate moment tensor solutions for the strongest earthquakes (3.1 ≤ M ≤ 4.0) We modeled waveforms of earthquakes recorded by at least 3 stations and at epicentral distances of 60-200 km. Due to the relatively small magnitudes of the chosen earthquakes, the waveforms used for the modeling were all pre-filtered by a high-pass filter of 0.01 Hz before the inversion procedure. During the inversion, we applied filters to the selected waveforms, ranging from 0.04 Hz to 0.12 Hz, according to the station epicentral distance. Green's functions are calculated by ISOLA using the discrete wavenumber method (Bouchon, 1981) according to the same local velocity model used during the location and relocation procedures. ISOLA solves for the best point source moment tensor by using the iterative deconvolution inversion algorithm of Kikuchi & Kanamori (1991) which seeks to minimize misfits between the observed and the modeled waveform for each earthquake. The robustness of the computed focal mechanisms is assessed by evaluating the quality metrics provided by ISOLA, mainly the variance reduction (VR) metric that controls the similarity between modeled and real waveforms, the condition number that controls the solution quality of the Green's functions matrix that result from the inversion as well as evaluating the Double-couple percentage of each solution. The latest edition of the software package includes a robust uncertainty assessment tool, computed from the covariance matrix of the source parameters and visualized through means of scatter plots of nodal lines, DC and CLVD percentage histograms etc. Since our dataset includes small magnitude earthquakes, a large number of our raw waveform data was too noisy, introducing challenges to the focal mechanism determination. In such cases, we utilized the CSPS (Cyclic Scanning of the Polarity Solutions) method, combining waveforms with good signal to noise ratio at close stations, with clear first motion polarities at many stations, regardless of distance (Fojtikova & Zahradnik, 2014; Zahradník et al., 2015). During CSPS, solutions are pre-constraint by polarities, providing a decisively non-unique focal mechanism. In the next step, ISOLA systematically scans the solutions that satisfy the observed polarities, choosing the focal mechanisms that provide the best waveform fit for the good quality stations, based on a user determined threshold of variance reduction. In cases when the CSPS method was used, we computed the candidate focal mechanisms using the FPFIT program (Reasenberg & Oppenheimer, 1985). Since most of the stations that recorded first motion polarities are very close to the earthquake sources, The focal parameters of the computed focal mechanisms are presented in Table 2.

Table 3: Focal parameters for the 17 computed focal mechanisms of the excitation

| Date | Time | Lat° | Lon° | Depth (Km) | Strike (°) | Dip (°) | Rake (°) | $M_w$ | ID |
|---|---|---|---|---|---|---|---|---|---|
| 03-03-2024 | 04:44:24 | 38.403 | 20.453 | 8.9 | 285 | 50 | 16 | 3.5 | 1 |
| 03-03-2024 | 10:36:13 | 38.393 | 20.479 | 7.9 | 100 | 44 | -1 | 3.6 | 2 |
| 03-03-2024 | 13:35:09 | 38.401 | 20.462 | 8.8 | 100 | 45 | 1 | 3.7 | 3 |
| 03-03-2024 | 20:33:11 | 38.398 | 20.483 | 8.8 | 96 | 62 | 6 | 3.5 | 4 |
| 03-03-2024 | 23:57:17 | 38.395 | 20.486 | 5.3 | 88 | 72 | 14 | 3.7 | 5 |
| 04-03-2024 | 07:06:44 | 38.395 | 20.483 | 9.5 | 284 | 73 | 6 | 4.0 | 6 |
| 04-03-2024 | 17:11:11 | 38.407 | 20.442 | 8.7 | 89 | 80 | -25 | 3.9 | 7 |
| 04-03-2024 | 19:18:46 | 38.401 | 20.484 | 13.9 | 285 | 88 | 30 | 3.7 | 8 |
| 05-03-2024 | 00:39:44 | 38.401 | 20.481 | 8.2 | 264 | 67 | 19 | 3.1 | 9 |
| 05-03-2024 | 17:17:56 | 38.399 | 20.475 | 8.6 | 127 | 65 | -2 | 3.2 | 10 |
| 05-03-2024 | 17:38:26 | 38.402 | 20.469 | 3.5 | 299 | 80 | -28 | 3.6 | 11 |
| 05-03-2024 | 17:48:31 | 38.396 | 20.483 | 8.7 | 304 | 79 | -21 | 3.1 | 12 |
| 05-03-2024 | 22:17:25 | 38.415 | 20.431 | 8.7 | 290 | 68 | 26 | 3.6 | 13 |
| 06-03-2024 | 00:07:24 | 38.404 | 20.456 | 8.9 | 109 | 55 | 28 | 3.4 | 14 |
| 06-03-2024 | 06:39:28 | 38.401 | 20.489 | 9.3 | 268 | 83 | 9 | 3.5 | 15 |
| 20-03-2024 | 00:30:01 | 38.392 | 20.497 | 7.1 | 141 | 81 | 7 | 3.3 | 16 |
| 11-04-2024 | 00:54:42 | 38.369 | 20.458 | 9.9 | 100 | 80 | 10 | 3.5 | 17 |

Figure 10 shows the focal mechanisms as equal area lower hemisphere projections, along with the first motion polarities used for their calculation with the CSPS method and the modeled waveforms for each solution. Filters applied to waveforms were chosen based on epicentral distance and noisiness. The utilization of clear first motion polarities for the computation provide stable and robust results, even in cases where the waveforms recorded for the studied earthquakes were noisy.

**Fig. 10** Computed focal mechanisms along with modeled waveforms. Greyed out waveforms were not used for the inversion. First motion polarities used for the CSPS method are marked on the solutions as **D** or **-** and **U** or **+** for Down and Up respectively. Polarities shown as "?" were unclear and therefore not used during computation. Pressure axis is shown as a blue P while Tension axis is shown as a yellow T on the computed solution

03 March 2024 04:44:24.30 (**Mw = 3.5**) **NP1**: 285° / 50° /16° , **NP2**: 185° /78° /139° , **Depth** = 8.9

03 March 2024 10:36:13 (**Mw**=3.6) **NP1:** 100°/ 44°/-1°, **NP2**: 191°/89°/-134°, **Depth** = 7.9

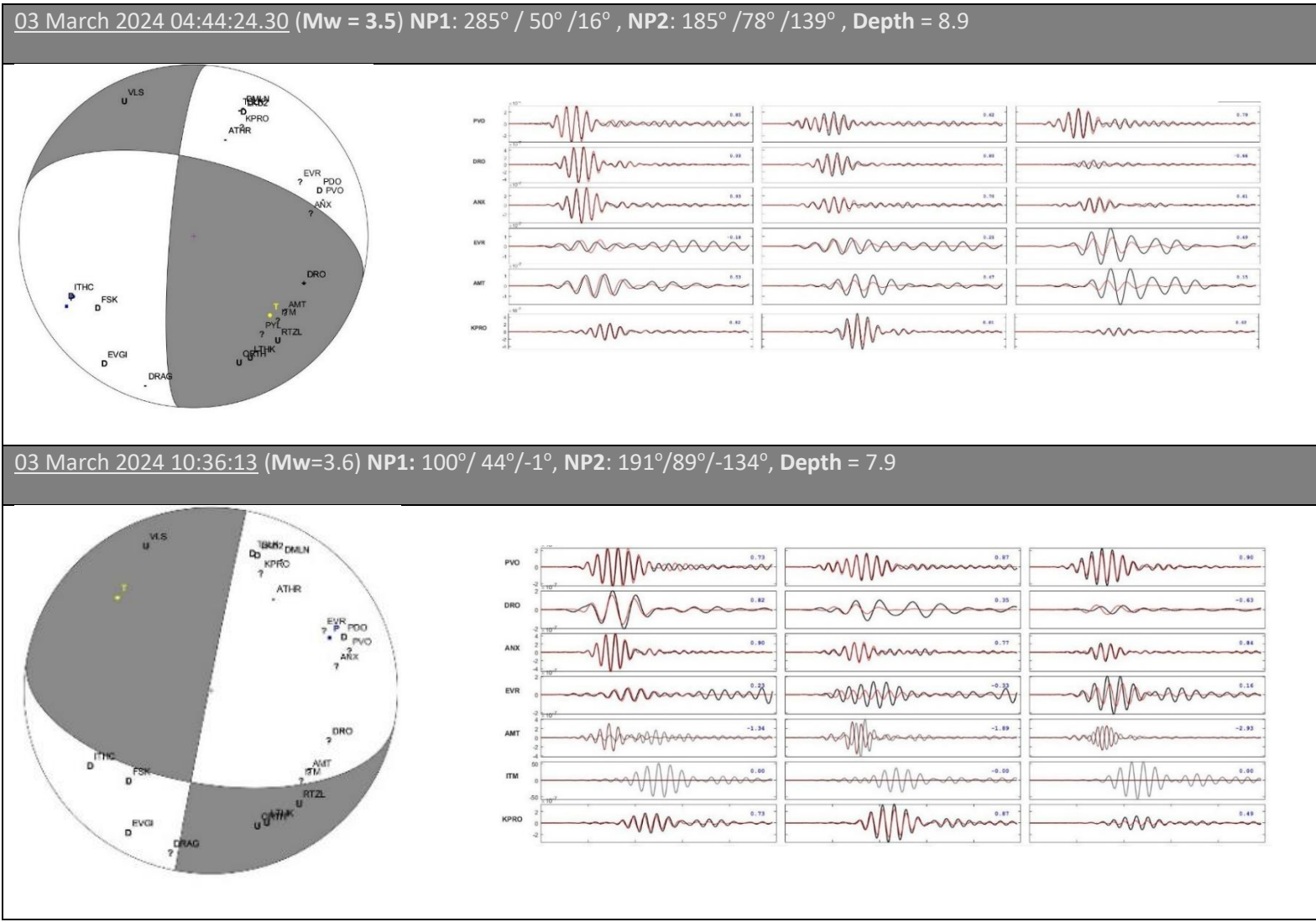

03 March 2024 13:35:09 (**Mw**=3.7) **NP1:** 100°/ 45°/1°, **NP2**: 9°/89°/135°, **Depth** = 8.8

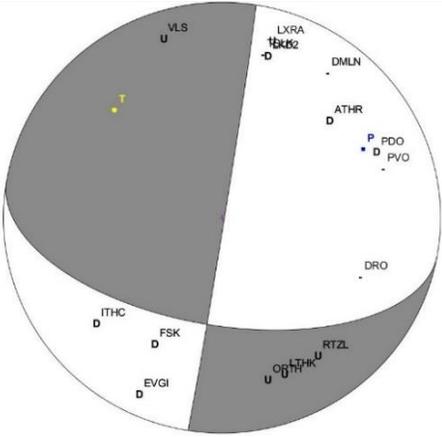 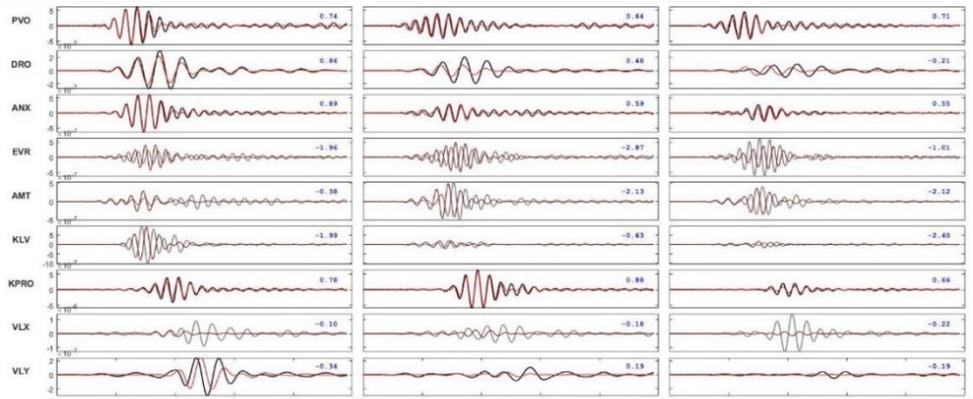

03 March 2024 20:33:11 (**Mw**=3.7) **NP1:** 96°/ 62°/6°, **NP2**: 3°/85°/151°, **Depth** = 8.8

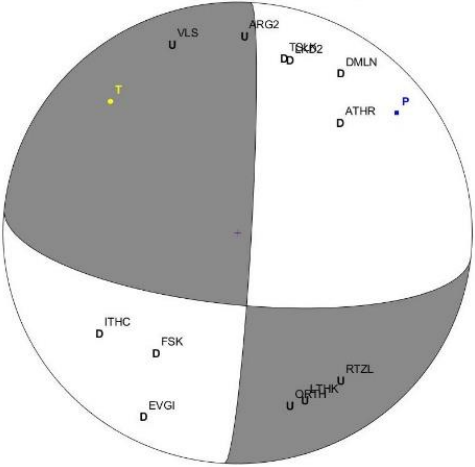 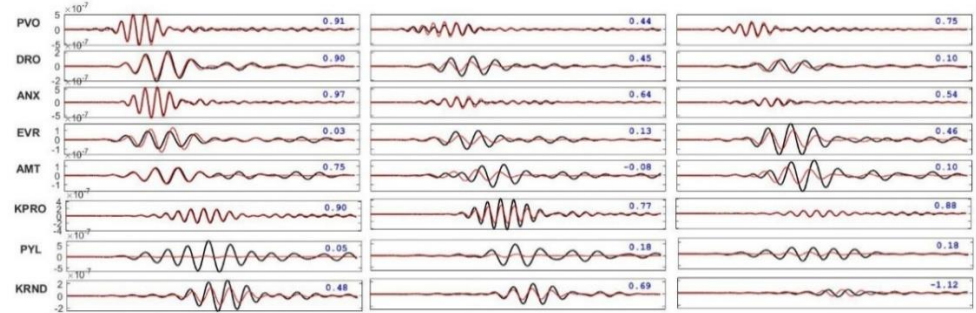

03 March 2024 23:57:17 (**Mw**=3.7) **NP1:** 88°/ 72°/14°, **NP2**: 354°/77°/161°, **Depth** = 5.3

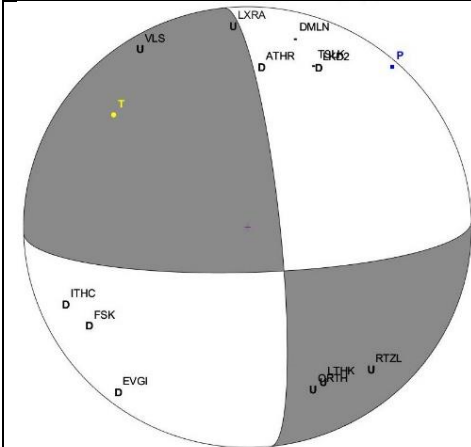 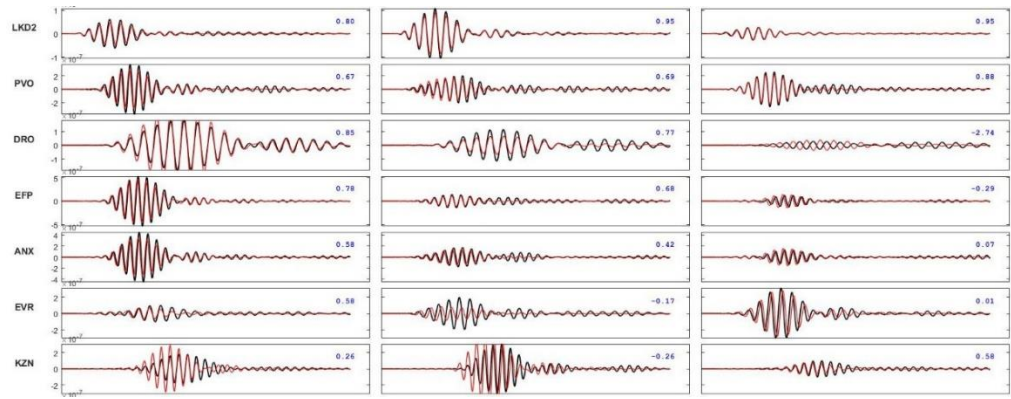

04 March 2024 07:06:44 (**Mw**=4.0) **NP1:** 284°/ 73°/6°, **NP2**: 192°/84°/163°, **Depth** = 9.5

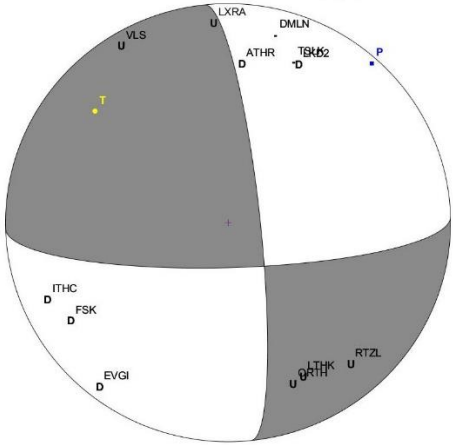
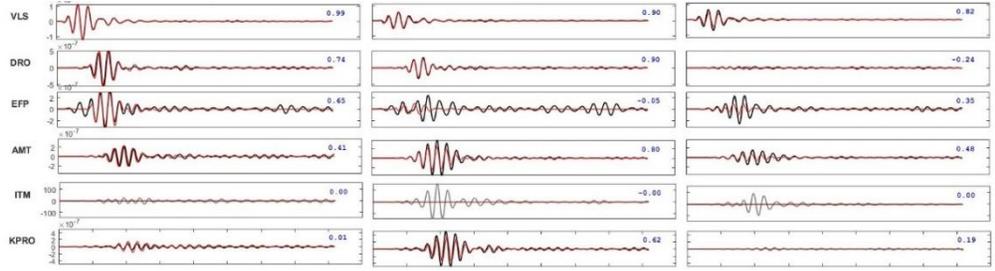

04 March 2024 17:11:11 (**Mw**=3.9) **NP1:** 89°/ 80°/-25°, **NP2**: 183°/66°/-169°, **Depth** = 8.7

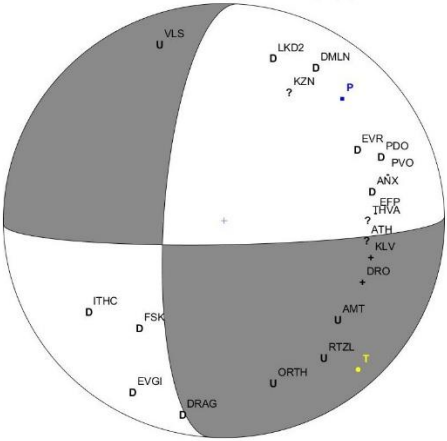
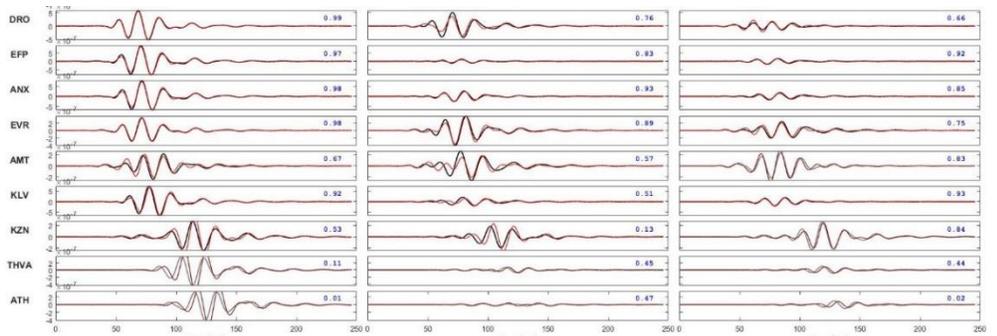

04 March 2024 19:18:46 (**Mw**=3.7) **NP1:** 285°/88°/30°, **NP2**: 194°/60°/177°, **Depth** = 13.9

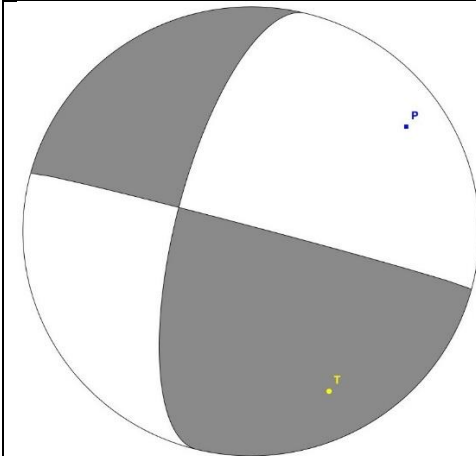
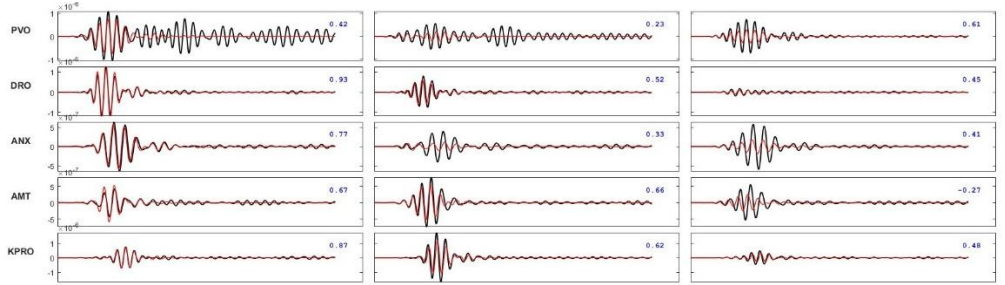

05 March 2024 00:39:44 (**Mw**=3.1) **NP1:** 264°/ 67°/19°, **NP2**: 167°/73°/156°, **Depth** = 8.2

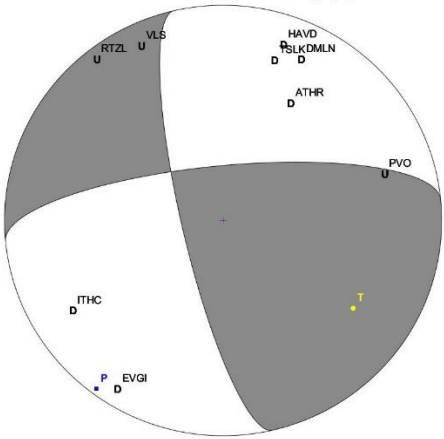 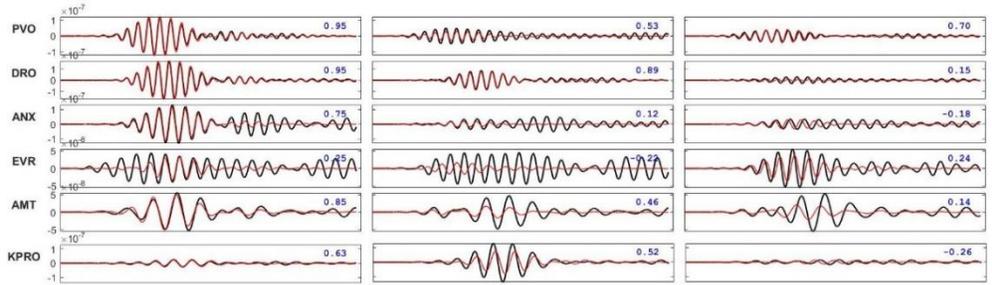

05 March 2024 17:17:56 (**Mw**=3.2) **NP1:** 127°/ 65°/-2°, **NP2**: 218°/89°/-155°, **Depth** = 8.6

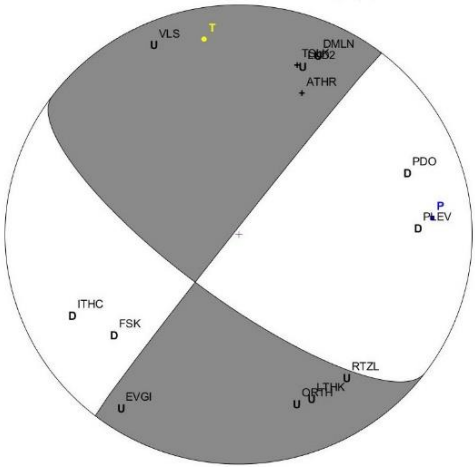 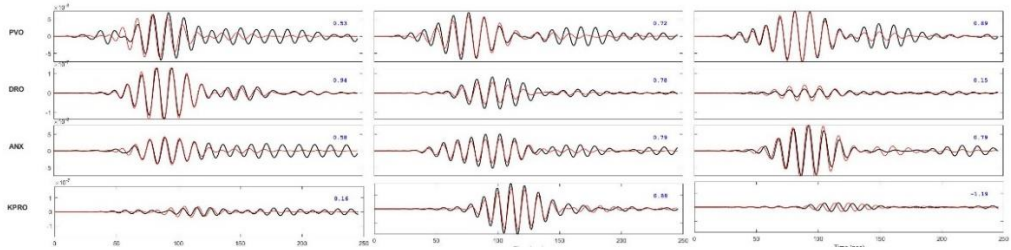

05 March 2024 17:38:26 (**Mw**=3.6) **NP1:** 299°/ 80°/-28°, **NP2**: 34°/62°/-169°, **Depth** = 3.5

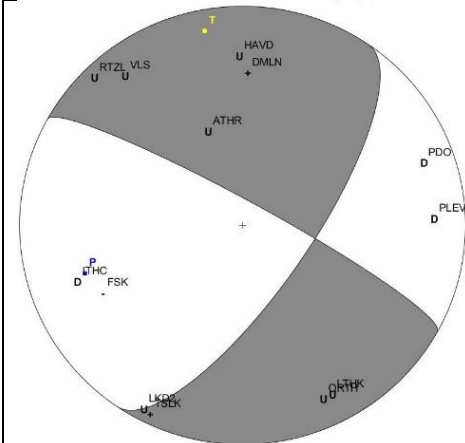 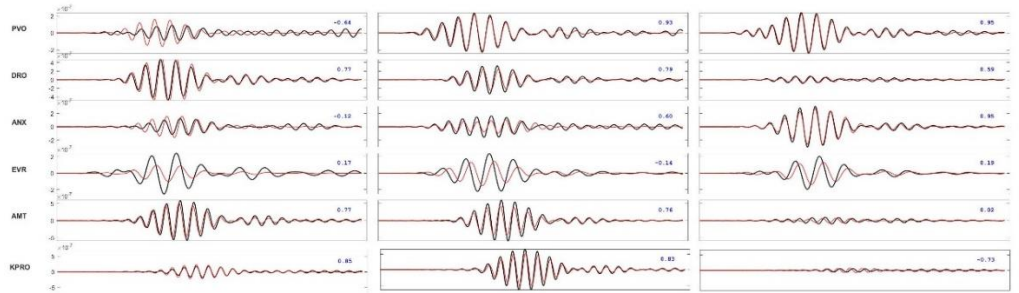

05 March 2024 17:48:26 (**Mw**=3.1) **NP1:** 304°/ 79°/-21°, **NP2**: 39°/69°/-168°, **Depth** = 3.5

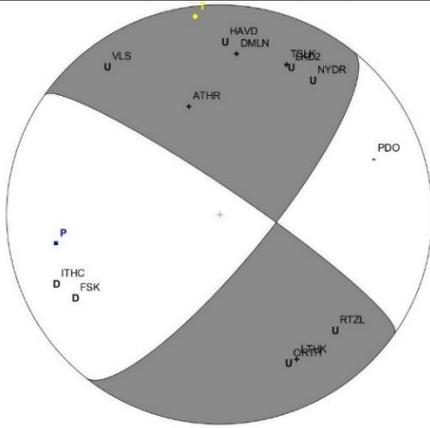 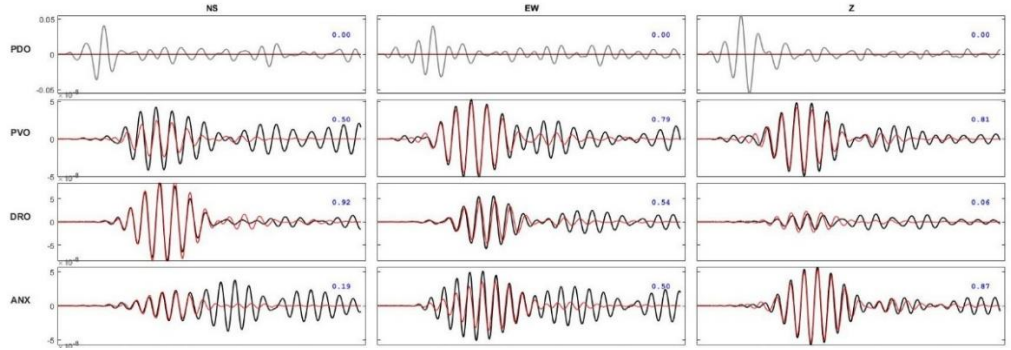

05 March 2024 22:17:25 (**Mw**=3.6) **NP1:** 290°/68°/26°, **NP2**: 190°/66°/155°, **Depth** = 8.7

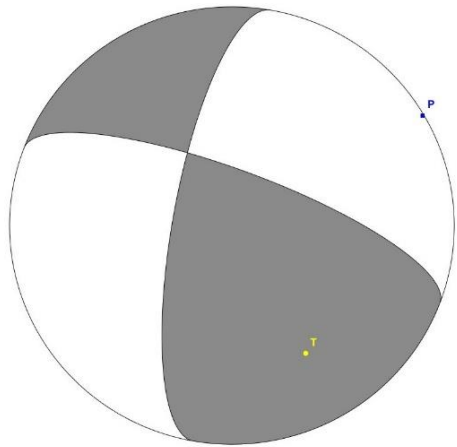 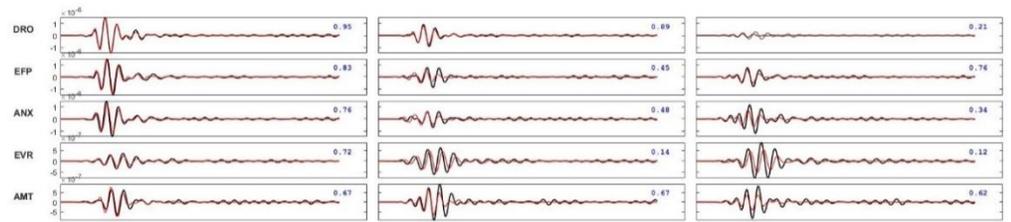

06 March 2024 00:07:24 (**Mw**=3.4) **NP1:** 109°/ 55°/28°, **NP2**: 2/67°/142°, **Depth** = 8.9

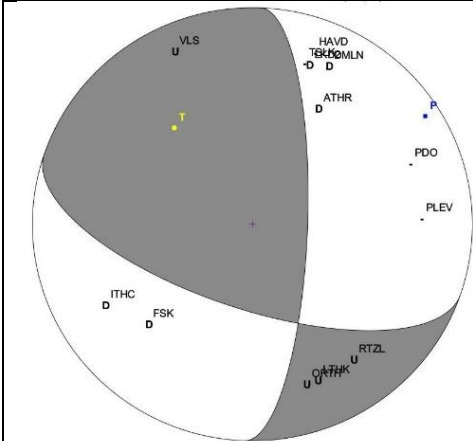 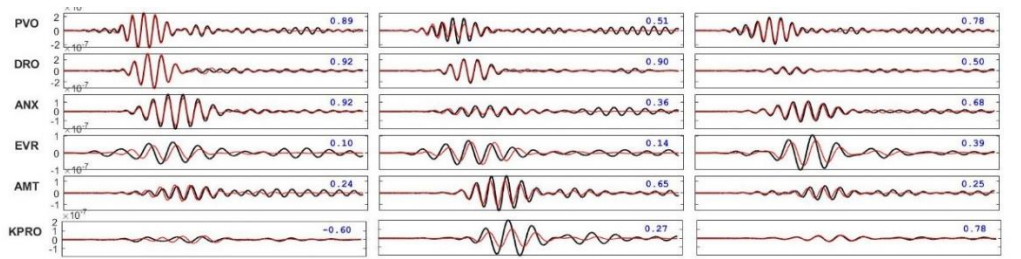

06 March 2024 06:39:28 (**Mw**=3.5) **NP1:** 268°/ 83°/9°, **NP2**: 177/81°/173°, **Depth** = 9.3

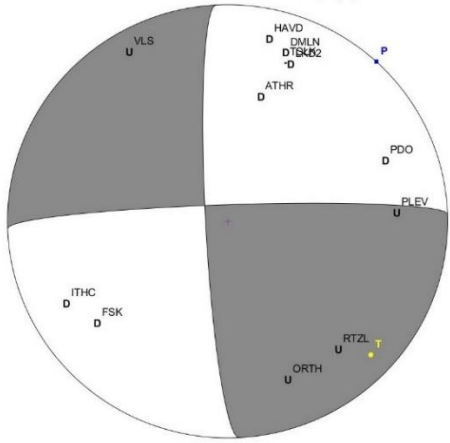
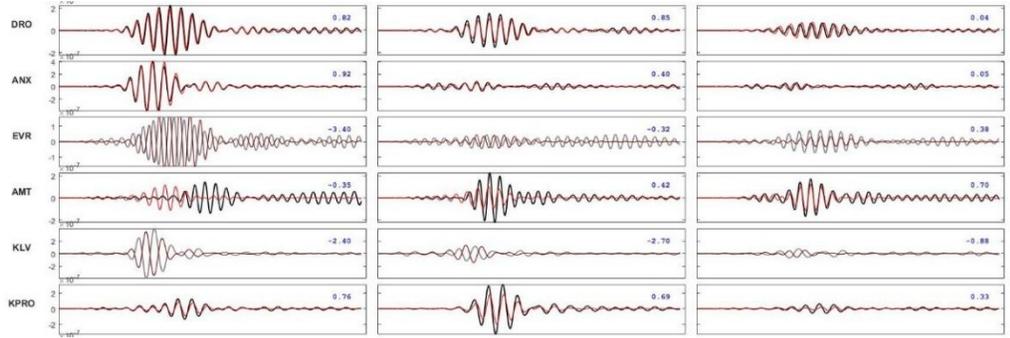

20 March 2024 00:30:01 (**Mw**=3.3) **NP1:** 141°/81°/7°, **NP2**: 50°/83°/170°, **Depth** = 7.1

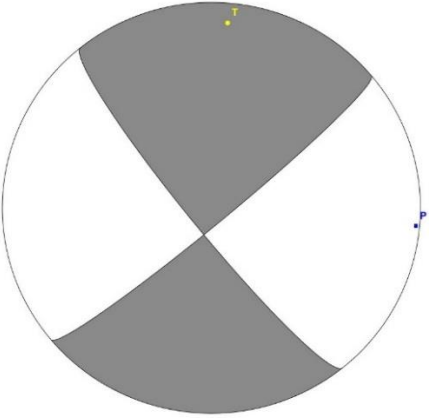
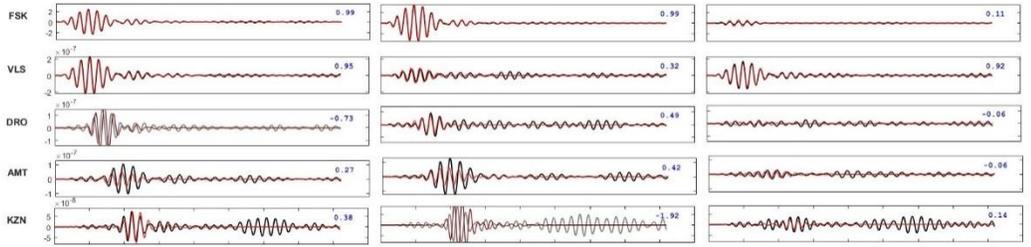

11 April 2024 00:54:42 (**Mw**=3.5) **NP1:** 100°/ 80°/10°, **NP2**: 8/80°/170°, **Depth** = 9.9

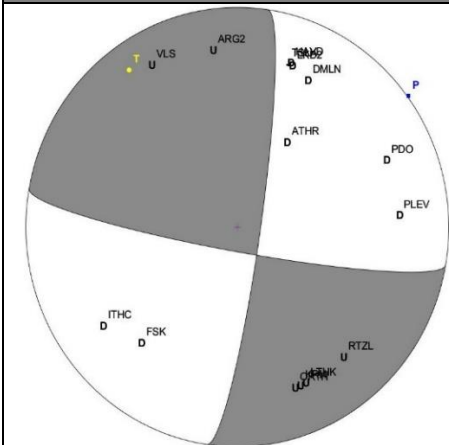
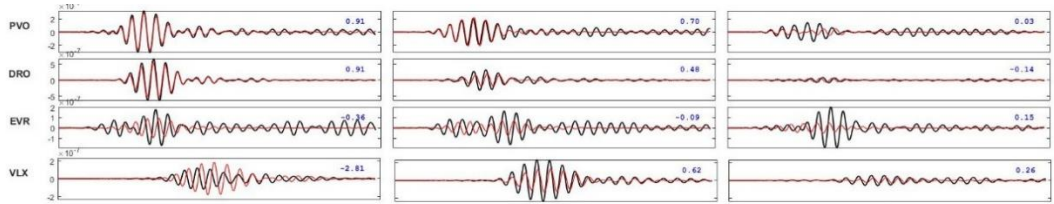

of Previous Strong Shocks. *Bulletin of the Seismological Society of America*, *92*(8), 3293–3308. https://doi.org/10.1785/0120000290

Papadimitriou, E., Karakostas, V., Mesimeri, M., Chouliaras, G., & Kourouklas, Ch. (2017). The Mw6.5 17 November 2015 Lefkada (Greece) Earthquake: Structural Interpretation by Means of the Aftershock Analysis. *Pure and Applied Geophysics*, *174*(10), 3869–3888. https://doi.org/10.1007/s00024-017-1601-3

Papazachos, B. C., & Comninakis, P. E. (1971). Geophysical and tectonic features of the Aegean Arc. *Journal of Geophysical Research*, *76*(35), 8517–8533. https://doi.org/10.1029/JB076i035p08517

Papazachos, B. C., Karakostas, V. G., Kiratzi, A. A., Margaris, B. N., Papazachos, C. B., & Scordilis, E. M. (2002). Uncertainties in the estimation of earthquake magnitudes in Greece. *Journal of Seismology*, *6*(4), 557–570. https://doi.org/10.1023/A:1021214126748

Papazachos, B. C., & Papazachou, C. (2003). *The earthquakes of Greece*. Ziti publications.

Papazachos, B. C., Scordilis, E. M., Panagiotopoulos, D. G., Papazachos, C. B., & Karakaisis, G. F. (2004). Global relations between seismic fault parameters and moment magnitude of earthquakes. *Bulletin of the Geological Society of Greece*, *36*(3), 1482. https://doi.org/10.12681/bgsg.16538

Peacock, D. C. P., Nixon, C. W., Rotevatn, A., Sanderson, D. J., & Zuluaga, L. F. (2016). Glossary of fault and other fracture networks. *Journal of Structural Geology*, *92*, 12–29. https://doi.org/10.1016/j.jsg.2016.09.008

Peng, Z., & Lei, X. (2024). Physical mechanisms of earthquake nucleation and foreshocks: Cascade triggering, aseismic slip, or fluid flows? *Earthquake Research Advances*, 100349. https://doi.org/10.1016/j.eqrea.2024.100349

Piombo, A., Martinelli, G., & Dragoni, M. (2005). Post-seismic fluid flow and Coulomb stress changes in a poroelastic medium. *Geophysical Journal International*, *162*(2), 507–515. https://doi.org/10.1111/j.1365-246X.2005.02673.x42                                    22/5/2025

# STATEMENTS AND DECLARATIONS

### Funding

This research is partially financially supported by the artEmis Project funded by the European Union, under Grant Agreement nr 101061712. The views and opinions expressed are, however, those of the author(s) only and do not necessarily reflect those of the European Union or European Commission–Euratom. Neither the European Union nor the granting authority can be held responsible for them.

### Competing Interests

The authors declare no competing interests.

# DATA AVAILABILITY

Waveform data used in this study are publicly available at https://doi.org/10.7914/SN/HT and at https://doi.org/10.7914/SN/HL. Last accessed through EIDA on 10 December 2024. Seismic catalog available upon request from corresponding author.